\documentclass[%
 reprint,
superscriptaddress,
 amsmath,amssymb,
 aps,
]{revtex4-2}
\usepackage{graphicx} % Required for inserting images
\usepackage{tikz}
\usepackage{natbib}%
\usepackage{hyperref}% add hypertext capabilities
\usepackage{subfigure}
\usepackage{amsmath}
\usepackage{comment}

\usepackage{svg}
\usepackage{booktabs}
\usepackage{multirow}
\usepackage{array}
\usepackage{makecell}

\begin{document}

\title{Electrically Tunable Interband Collective Excitations in Biased Bilayer and Trilayer Graphene}

\author{Tomer Eini}
\affiliation{School of Electrical Engineering, Faculty of Engineering, Tel Aviv University, Tel Aviv 6997801, Israel}
\author{M. F. C. Martins Quintela}
\affiliation{Department of Physics and Physics Center of Minho and Porto Universities (CF–UM–UP), Campus of Gualtar, 4710-057, Braga, Portugal}
\affiliation{International Iberian Nanotechnology Laboratory (INL), Av. Mestre José Veiga, 4715-330, Braga, Portugal}
\affiliation{Department of Materials and Production, Aalborg University, 9220 Aalborg Øst, Denmark}
\affiliation{Departamento de Qu\'{i}mica, Universidad Aut\'{o}noma de Madrid, 28049 Madrid, Spain}
\affiliation{Condensed Matter Physics Center (IFIMAC), Universidad Aut\'{o}noma de Madrid, 28049, Madrid, Spain}
\author{J. C. G. Henriques}
\affiliation{International Iberian Nanotechnology Laboratory (INL), Av. Mestre José Veiga, 4715-330, Braga, Portugal}
\affiliation{Universidade de Santiago de Compostela, 15782 Santiago de Compostela, Spain}
\author{R. M. Ribeiro}
\affiliation{International Iberian Nanotechnology Laboratory (INL), Av. Mestre José Veiga, 4715-330, Braga, Portugal}
\affiliation{Centro de Física das Universidades do Minho e do Porto (CF-UM-UP) e Departamento de Física, Universidade do Minho, P-4710-057 Braga, Portugal}
\author{Yarden Mazor}
\affiliation{School of Electrical Engineering, Faculty of Engineering, Tel Aviv University, Tel Aviv 6997801, Israel}
\author{N. M. R. Peres}
\affiliation{International Iberian Nanotechnology Laboratory (INL), Av. Mestre José Veiga, 4715-330, Braga, Portugal}
\affiliation{Centro de Física das Universidades do Minho e do Porto (CF-UM-UP) e Departamento de Física, Universidade do Minho, P-4710-057 Braga, Portugal}
\affiliation{POLIMA—Center for Polariton-driven Light-Matter Interactions, University of Southern Denmark, Campusvej 55, DK-5230 Odense M, Denmark}
\author{Itai Epstein}
\email{itaieps@tauex.tau.ac.il}
\affiliation{School of Electrical Engineering, Faculty of Engineering, Tel Aviv University, Tel Aviv 6997801, Israel}
\affiliation{Center for Light-Matter Interaction, Tel Aviv University, Tel Aviv 6997801, Israel}
\affiliation{QuanTAU, Quantum Science and Technology Center, Tel Aviv University, Tel Aviv 6997801, Israel}

\date{December 2024}

\begin{abstract}
Collective excitations of charged particles under the influence of an electromagnetic field give rise to a rich variety of hybrid light-matter quasiparticles with unique properties. In metals, intraband collective response manifested by negative permittivity leads to plasmon-polaritons with extreme field confinement, wavelength "squeezing", and potentially low propagation losses. In contrast, photons in semiconductors commonly couple to interband collective response in the form of exciton-polaritons, which give rise to completely different polaritonic properties, described by a superposition of the photon and exciton and an anti-crossing of the eigenstates. In this work, we identify the existence of plasmon-like collective excitations originating from the interband excitonic response of biased bilayer and trilayer graphene, in the form of graphene-exciton-polaritons (GEPs). We find that GEPs possess electrically tunable polaritonic properties and discover that such excitations follow a universal dispersion law for all surface polaritons in 2D excitonic systems. Accounting for nonlocal corrections to the excitonic response, we find that the GEPs exhibit confinement factors that can exceed those of graphene plasmons, and with moderate losses. These predictions of plasmon-like interband collective excitations in biased graphene systems open up new research avenues for tunable polaritonic phenomena based on excitonic systems, and the ability to control and manipulate such phenomena at the atomic scale. 
\end{abstract}

\maketitle

\section{Introduction}
Collective excitation can be described as the interaction between dipole oscillations of charged (quasi-) particles and the electromagnetic field oscillations of an impinging photon. Fundamentally, this interaction forms a new quasi-particle, i.e. the polariton, which inherits its properties from both the photon and the material excitation, constituting an eigenmode of the overall electromagnetic field in the system. The vast variety of such interacting quasi-particles give rise to numerous types of polaritons, such as plasmon-polaritons, exciton-polaritons, phonon-polaritons, hyperbolic-polaritons, cooper pair polaritons, magnon-polaritons, and more \cite{Basov2020PolaritonPanorama, Basov2016PolaritonsMaterials, Low2016PolaritonsMaterials, Jablan2009PlasmonicsFrequencies, Koppens2011,  Dai2014TunableNitride, Caldwell2015Low-lossPolaritons, Caldwell2014Sub-diffractionalNitride, Schneider2018Two-dimensionalCoupling}.  

Different types of polaritons can be classified to families with similar properties. For example, surface polaritons that propagate in the in-plane direction of an interface between two materials, where one of the materials exhibits a negative real part of its permittivity, are characterized as highly confined optical modes due to their ability to carry large momentum in the propagation direction. Surface polaritons can thus be supported in any material exhibiting negative permittivity, commonly stemming from the intraband material response, e.g. surface-plasmon-polaritons in metals \cite{Maier2007Plasmonics:Applications, Barnes2003SurfaceOptics, Ebbesen1998ExtraordinaryArrays, Epstein2016Surface-plasmonHolography}, surface-exciton-polaritons in semiconductors \cite{Agranovich2014SurfacePolaritons, Lagois1976ExperimentalPolaritons, Lagois1978DispersionPolaritons, Elrafei2024GuidingSuperlattices, Tokura1982SurfaceZnO}, graphene plasmons \cite{Koppens2011, Jablan2009PlasmonicsFrequencies, Chen2012OpticalPlasmons, Fei2012Gate-tuningNano-imaging, Hesp2021ObservationGraphene}, surface and hyperbolic phonon-polaritons in polar dielectric \cite{Caldwell2015Low-lossPolaritons, Caldwell2014Sub-diffractionalNitride, Dai2014TunableNitride, Hillenbrand2002Phonon-enhancedScale, Caldwell2013Low-lossResonators}, and more. Semiconductors, on the other hand, which are commonly described by an interband optical response, give rise to exciton-polaritons with different polaritonic properties. Stemming from the interaction of a semiconductor exciton with a cavity photon, such exciton-polaritons are described as a superposition of the exciton and the photon, resulting in splitting and an anti-crossing of the new eigenstates \cite{Weisbuch1992ObservationMicrocavity, Kavokin2010Exciton-polaritonsPerspectives, Deng2010Exciton-polaritonCondensation, Timofeev2012ExcitonFrontiers, Yamamoto1999MesoscopicOptics, Liu2014StrongCrystals, Schneider2018Two-dimensionalCoupling}.

In the last decade, two-dimensional (2D) materials have drawn significant interest as a versatile platform for studying polaritons at the atomic scale \cite{Basov2020PolaritonPanorama, Basov2016PolaritonsMaterials, Low2016PolaritonsMaterials, Caldwell2015Low-lossPolaritons}. The unique properties of these materials, such as their low dimensionallity, pronounced quantum effects, and natural anisotropy, give rise to quasi-particles that strongly interact with light, which can be readily controlled via their surrounding, charged carrier denisty, and temperature \cite{Chen2012OpticalPlasmons, Fei2012Gate-tuningNano-imaging, Alonso-Gonzalez2016AcousticNanoscopy, Alonso-Gonzalez2014ControllingPatterns, Ju2011GrapheneMetamaterials, Ni2018FundamentalPlasmonics}. Moreover, they can be easily integrated into van der Waals heterostructures and cavities yielding enhanced response and  hybridizatoin \cite{Schneider2018Two-dimensionalCoupling, Liu2014StrongCrystals, Sun2017OpticalPolaritons, Epstein2020Near-UnityCavityb, Horng2020PerfectCrystal, Fang2019ControlHeterostructures, Horng2019EngineeringSemiconductors}. 

Recently, several graphene-based material systems have been demonstrated to exhibit unique excitonic response, such as Bernel-stacked and twisted Bilayer Graphene (BLG), and Rhombohedral Trilayer Graphene (RTG) \cite{Henriques2022AbsorptionGraphene, Quintela2022TunableGraphene, Ju2017TunableGraphene, Park2010TunableGraphene, Ju2020UnconventionalGraphene, Duarte2024MoirePressure}. In their native form, the BLG and RTG are semi-metallic with a zero bandgap and a non-conical band structure. However, when encapsulated in an isolating material, such as hexagonal-boron-nitride (hBN), inversion symmetry can be broken by applying a bias that forms an electric field across the graphene layers. This results in the opening of a bandgap and the system thus becomes semiconducting \cite{Castro2007BiasedEffect, Zhang2009DirectGraphene., Oostinga2007Gate-inducedDevices}, with the ability to support excitons with large binding energies \cite{Henriques2022AbsorptionGraphene, Quintela2022TunableGraphene, Park2010TunableGraphene, Ju2017TunableGraphene, Ju2020UnconventionalGraphene}, similar to excitons in 2D semiconductors \cite{Wang2018Colloquium:Dichalcogenides}. 

In this work, we unveil the existence of plasmon-like collective excitations, typically associated with intraband responses, within the interband excitonic regime of biased BLG and RTG (BBLG and BRTG), in the form of graphene-exciton-polaritons (GEPs). We find that GEPs possess electrically tunable polaritonic properties at far-infrared (FIR) frequencies, and we discover that such interband excitation follow a universal dispersion law for all surface polaritons in 2D excitonic systems. Taking into account the appropriate nonlocal corrections, we analytically derive the dispersion relations of the GEPs and study their electrically tunable polaritonic properties: confinement, loss, and field distribution. We find that GEPs can reach confinement factors that are larger than those of graphene plasmons, and with moderate losses that would enable their observation in cryo-SNOM experiments. Furthermore, we show that the GEPs' properties can be electrically controlled and spectrally tuned through the biasing of the BBLG and BRTG systems.

\section{Optical excitonic conductivity of BBLG and BRTG}
\label{section cond}

\begin{figure}[t]
    \centering
    \hspace*{-13pt} % Shift the figure slightly to the left
    \begin{tikzpicture}
    \node[anchor=south west,inner sep=0] (image) at (0,0) {\includegraphics[width=1.03\columnwidth]{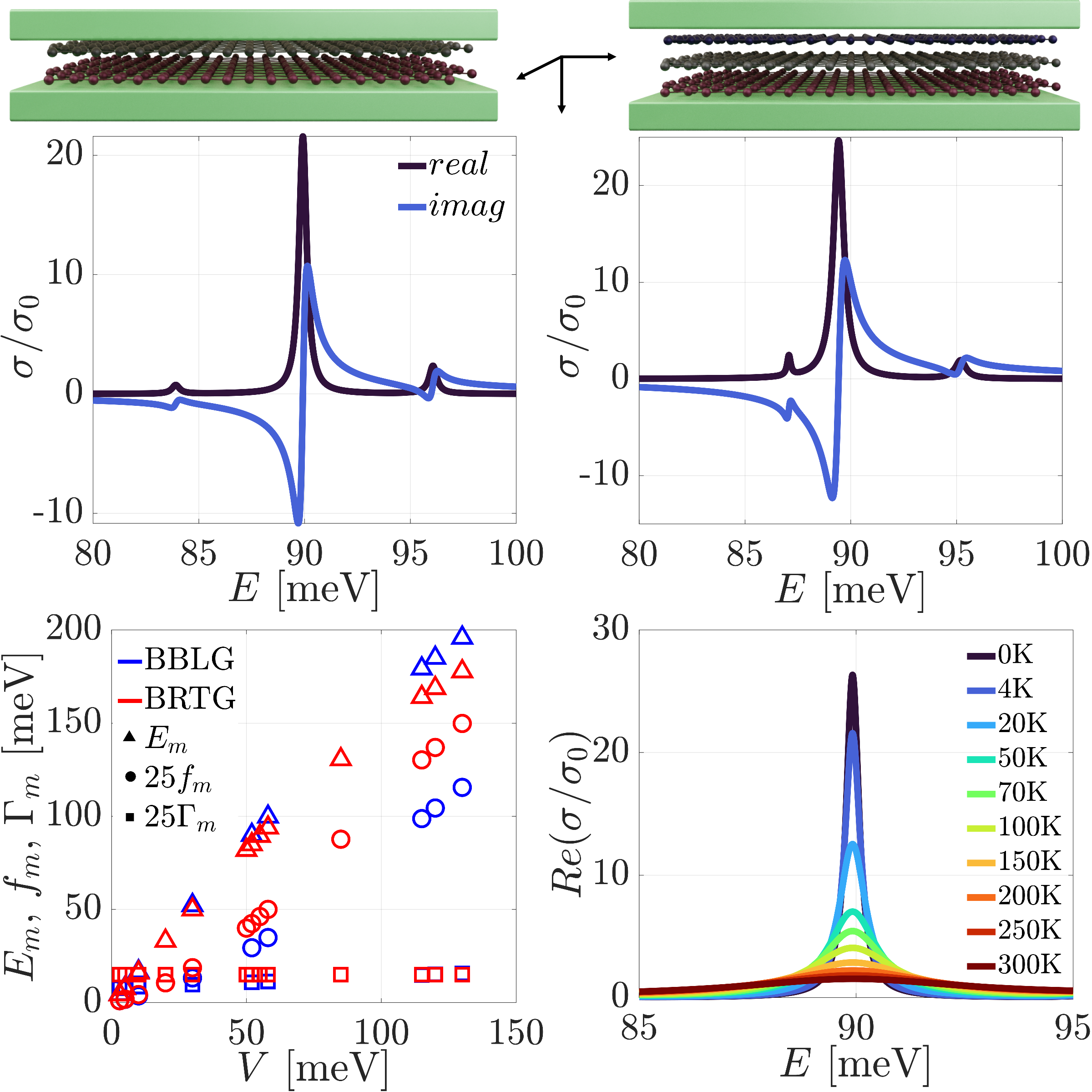}};
    %\label{''}
    \begin{scope}[x={(image.south east)},y={(image.north west)}]
    \node at (0.01,0.845) {(a)};
    \node at (0.503,0.845) {(b)};
    \node at (0.01,0.43) {(c)};
    \node at (0.503,0.43) {(d)};
    \node at (0.51,0.885) {$z$};
    \node at (0.575,0.945) {$x$};
    \node at (0.465,0.925) {$y$};
    \node at (0.21,0.85) {$n=m$};
    \node at (0.69,0.84) {$n=m$};
    \end{scope}
    \draw[->, thick, red] (6.2, 7.4) -- (6.5, 7);
    \draw[->, thick, red] (2.05, 7.4) -- (2.25, 7);
    \end{tikzpicture}
    \caption{Optical conductivity of (a) BBLG for $V=52\,\mathrm{meV}$, and (b) BRTG for $V=55\,\mathrm{meV}$, obtained from Eq. \ref{Eq conductivity}, both for $T=4\,\mathrm{K}$.
    Both the BBLG and BRTG exhibit positive values of the imaginary part of the conductivity, due to the excitonic resonances. The main resonances $n=m$ are marked with arrows.
    The configurations of BBLG and BRTG encapsulated by hBN are illustrated above the figure.
    (c) Voltage dependence of the exciton energy, oscillator strength and non-radiative decay rate of the main resonance of BBLG (blue) and BRTG (red) for $T=4\,\mathrm{K}$.
    (d) Temperature dependence of the real part of the conductivity of BBLG for $V=52\,\mathrm{meV}$.}
    \label{fig cond}
\end{figure}
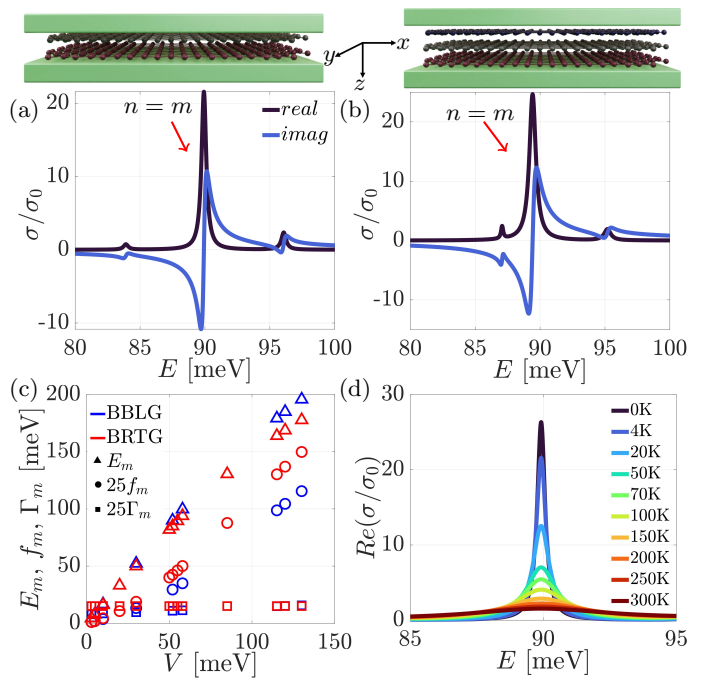

 The systems under investigation in this work are the BBLG and the BRTG, each separately encapsulated by hBN (Fig. \ref{fig cond}). The gap is opened by an applied bias between two electrodes placed above and below the encapsulating hBN, forming a dual-gated metal-oxide-semiconductor-field-effect-transistor (MOSFET), with the electric field across the graphene system being electrically controlled by the bias \cite{Henriques2022AbsorptionGraphene, Quintela2022TunableGraphene, Park2010TunableGraphene, Ju2017TunableGraphene, Ju2020UnconventionalGraphene}.

Under these conditions, the optical response of the BBLG and BRTG systems is given by the inclusion of  the excitonic response into their optical conductivities, taking the form \cite{Henriques2022AbsorptionGraphene,Quintela2022TunableGraphene}:
\begin{equation} \label{Eq conductivity}
    \sigma=4i\sigma_0\sum_{n} \frac{f_n}{E -E_n+i\frac{\Gamma_n}{2}},
\end{equation}
where $\sigma_0=\frac{e^2}{4\hbar}$ is the
universal conductivity of graphene, $E$ is the energy, $f_n$, $E_n$ and $\Gamma_n$ are the oscillator strength, exciton energy, and non-radiative decay rate of the excitonic resonance of the $n$th exciton Rydberg series.
This optical conductivity is obtained directly from the dynamical equation of the density matrix when considering a perturbation by a monochromatic field, taking into account the effect of the applied bias \cite{Pedersen2015IntrabandGeneration,Aversa1995NonlinearAnalysis,E-PeriodicaElectrodynamics, Lehmann1954UberFelder,Henriques2022AbsorptionGraphene,Quintela2022TunableGraphene, Quintela2024TunableGraphene} (see SI section A%appendix \ref{Section app cond}
).
From now on we will focus on the main resonance $n=m$, with the largest oscillator strength, as it dominates the optical response (Fig. \ref{fig cond} (a) and (b)).

The conductivities of BBLG and BRTG from Eq. \ref{Eq conductivity} are presented in Fig. \ref{fig cond} (a) and (b), respectively. It can be seen that in both cases, positive values of the imaginary part of $\sigma$ can be observed, stemming from the excitonic resonances of the biased graphene systems, as will be discussed further in section \ref{Section_GEP}. In addition, we note that these conductivities depend on the bias voltage \cite{Henriques2022AbsorptionGraphene, Quintela2022TunableGraphene} and temperature \cite{Epstein2020HighlySemiconductors}, affecting the oscillator strength and the non-radiative decay rate (Fig. \ref{fig cond} (c) and (d)) (see SI sections B and C%Appendices \ref{section app vol and temp cond} and \ref{section app temperature}
). In Fig. \ref{fig cond} (d), the affect of temperature on the BBLG conductivity can be seen, resulting in the broadening of the excitonic resonance due to interaction with phonons. The exact temperature dependence of the BRTG conductivity could not be similarly calculated, owing to the lack of information in the literature on the parameters required for the phonon-exciton scattering computation (SI section C%appendix \ref{section app temperature}
). Thus, following \cite{Quintela2022TunableGraphene} we take the BRTG linewidth to be constant.

\section{Graphene-Exciton-Polariton} \label{Section_GEP}

The basis of the existence of surface polaritons and their unique properties stems from materials possessing negative real part of their permittivity (positive values of the imaginary part of their conductivity) \cite{Maier2007Plasmonics:Applications, Dai2014TunableNitride, Epstein2020HighlySemiconductors, Eini2022Valley-polarizedFrequencies, Li2015HyperbolicFocusing, Kats20242DMaterials}. Therefore, the positive imaginary values exhibited by the BBLG and BRTG conductivities (Fig \ref{fig cond} (a) and (b)) imply that these systems support surface polaritons. Since it is the excitonic contribution to the conductivity of the graphene systems that provides these positive values in the imaginary part, and thus its polaritonic response, we refer to these as Graphene-Exciton-Polaritons (GEPs).

The dispersion relation of the GEPs in these structures is described by (SI section E%appendix \ref{Section app GEP}
):

\begin{equation}
\label{Eq DR_GEP}        q_{GEP}=\frac{2i\varepsilon_0\varepsilon_{hBN_{eff}}\omega}{\sigma},
\end{equation}
where $q_{GEP}$ is the component of the wavevector in the
$x$ direction (the momentum), $\omega$ is the angular frequency, $\varepsilon_{hBN_{xx}}$, $\varepsilon_{hBN_{xx}}$ and $\varepsilon_{hBN_{eff}}=\sqrt{\varepsilon_{hBN_{xx}}\varepsilon_{hBN_{zz}}}$ are the in-plane $x$, out-of-plane $z$ and effective permittivity of hBN \cite{Woessner2014HighlyHeterostructures}, and $\sigma$ is the conductivity of either BBLG or BRTG (Eq. \ref{Eq conductivity}). 

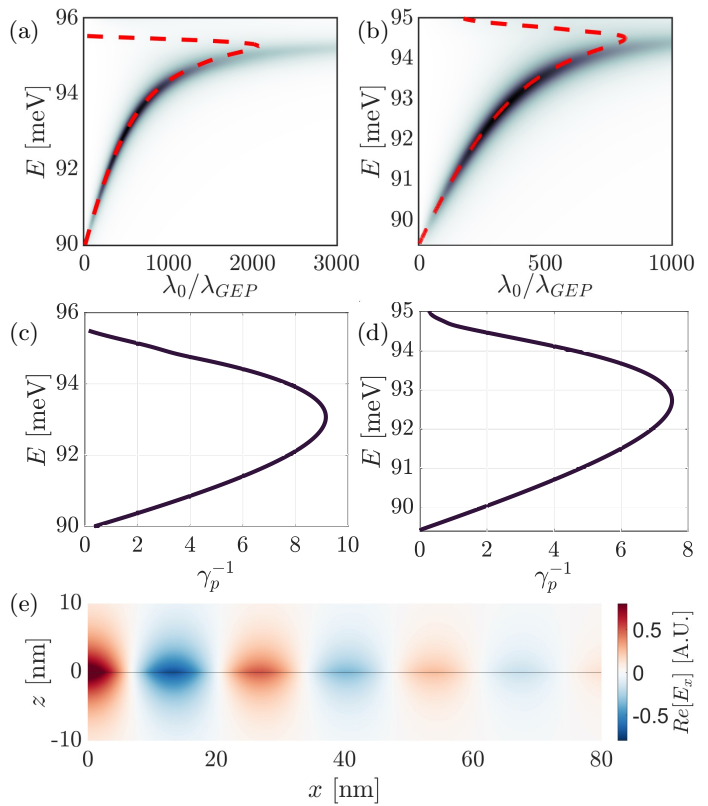
\begin{figure} [t]
    \centering
    \hspace*{-13pt} % Shift the figure slightly to the left
    \begin{tikzpicture}
    \node[anchor=south west,inner sep=0] (image) 
    at (0,0) 
    {\includegraphics[width=1.03\columnwidth]{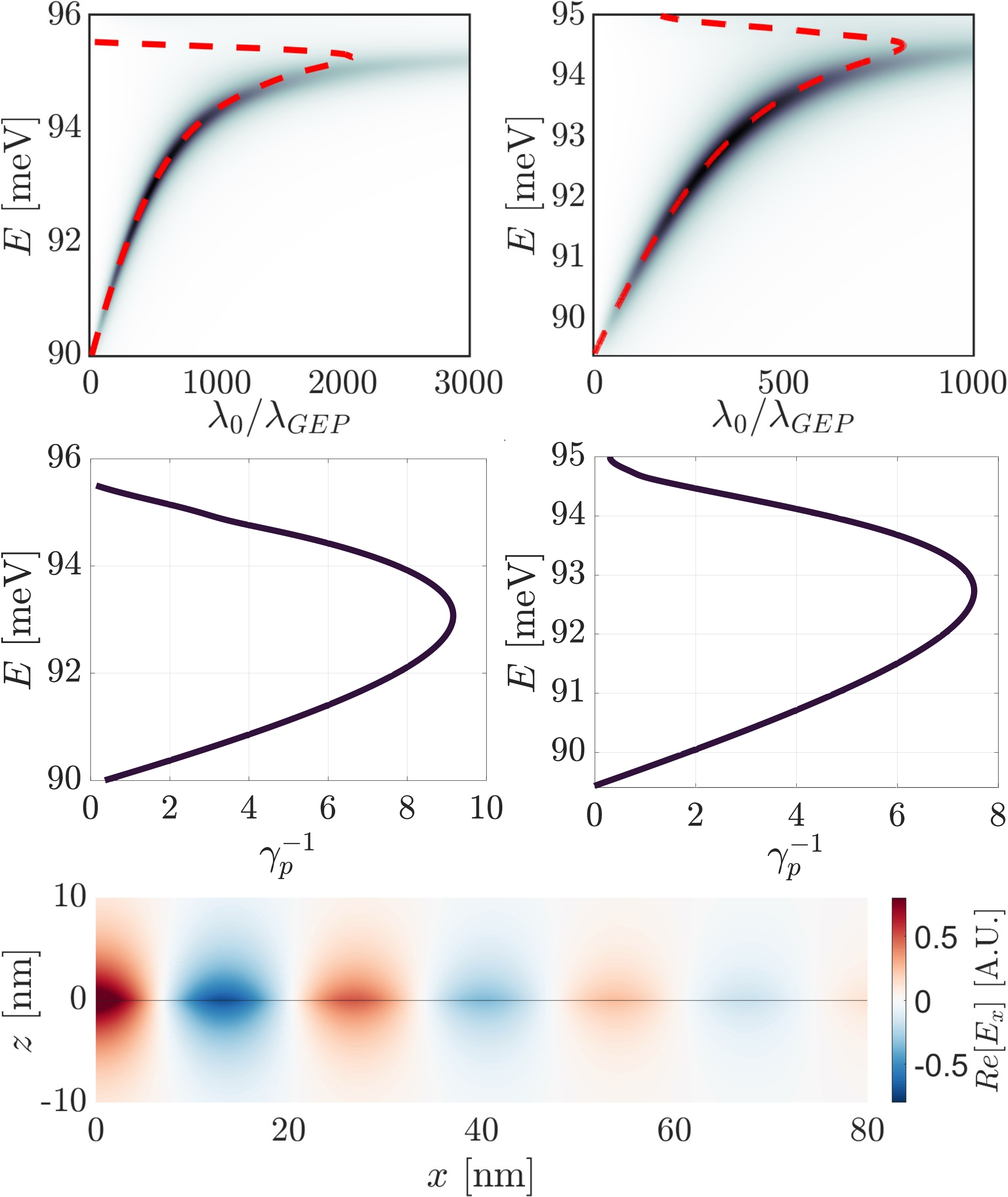}};
    %\label{''}
    \begin{scope}[x={(image.south east)},y={(image.north west)}]
    \node at (0,0.97) {(a)};
    \node at (0.52,0.97) {(b)};
    \node at (0,0.59) {(c)};
    \node at (0.52,0.59) {(d)};
    \node at (0,0.25) {(e)};
    \end{scope}
    \end{tikzpicture}
    \caption{Polaritonic properties of GEP.
    Confinement factor of GEP $\lambda_0 / \lambda_{GEP}$, for BBLG for $V=52\,\mathrm{meV}$ (a) and for BRTG for $V=55\,\mathrm{meV}$ (b), both for $T=4\,\mathrm{K}$, calculated from Eq. \ref{Eq DR_GEP} (dashed red line) and simulated using a TMM (colormap).
    The confinement reaches two and three orders of magnitude.
    Inverse damping ratio of GEP $\gamma_p^{-1}=Re(q_{GEP}) / Im(q_{GEP})$, for BBLG (c) and BRTG (d). (e) The electromagnetic field distribution of GEP in BBLG at $E=93\,\mathrm{meV}$, the graphene is marked at $z=0$.
    The polaritons propagate with losses in the in-plane direction and decay in the out-of-plane direction.}
    \label{fig_DR}
\end{figure}

The general properties of the GEPs can now be assessed by evaluating the confinement factor, $\lambda_0 / \lambda_{GEP}= Re(q_{GEP}) / k_0$, and the loss figure of merit described by inverse damping ratio $\gamma_p^{-1}=Re(q_{GEP}) / Im(q_{GEP})$ \cite{Woessner2014HighlyHeterostructures, Caldwell2015Low-lossPolaritons, Ni2018FundamentalPlasmonics}, where $\lambda_0=\frac{2\pi}{k_0}$ is the free-space wavelength, $k_0=\frac{\omega}{c}$ is the free-space wavenumber, $c$ is the speed of light in vacuum, and $\lambda_{GEP}=\frac{2 \pi}{Re(q_{GEP})}$ is the GEP wavelength. Figure \ref{fig_DR} (a) and (c) show the confinement factor and inverse damping ratio of BBLG GEPs, respectively, for the optimal voltage $V=52\,\mathrm{meV}$ and temperature $T=4\,\mathrm{K}$,  i.e., the largest confinement factors with minimal propagation losses. 
Excellent agreement between the transfer matrix method (TMM) simulation (colormap) and the analytical solution obtained from Eq. \ref{Eq DR_GEP} (dashed red line) can be seen.
Similarly, Fig. \ref{fig_DR} (b) and (d) show the confinement factor and inverse damping ratio of BRTG GEPs, respectively, for the optimal voltage $V=55\,\mathrm{meV}$ for BRTG, and temperature $T=4\,\mathrm{K}$. It can be seen from Fig. \ref{fig_DR} (a) and (b) that the GEPs in both systems are extremely confined, showing up to two orders of magnitude factors for the BRTG and even three orders of magnitude for the BBLG. These confinement factors are much larger than those observed for graphene plasmons \cite{Alonso-Gonzalez2016AcousticNanoscopy, Lundeberg2017TuningPlasmonics, Iranzo2018ProbingHeterostructure, Epstein2020Far-fieldVolumes, Menabde2021Real-spaceDeposition}. However, we note that these confinement factors are accompanied by propagation lengths over ten times smaller than those of graphene plasmons at low temperatures \cite{Ni2018FundamentalPlasmonics}. In addition, the results obtained in Fig. \ref{fig_DR} are based on the local description of the optical response, which only applies for low momentum modes. For the appropriate description of high momentum modes, one must take into account the nonlocal, momentum-dependent response, which will alter both the confinement factor and propagation losses of the graphene systems, as will be addressed in section \ref{section nonlocal}. For completeness, Fig. \ref{fig_DR} (e) show the electric field distribution of the BBLG GEPs (SI section E%appendix \ref{Section app GEP}
).

Intuitively, we can obtain simpler expressions for the confinement factor and inverse damping ratio by examining the conductivity near this main resonance, where $|E-E_n| \gg \Gamma_n$ for $n \neq m$ and $|E-E_m|\lesssim\Gamma_m$, and thus the contribution of additional resonances is negligible. We can therefore approximate the conductivity as:
\begin{equation} 
\label{Eq conductivity apprx}
    \sigma \approx \frac{4i\sigma_0f_m}{E-E_m+i\frac{\Gamma_m}{2}}.
\end{equation}

By inserting Eq. \ref{Eq conductivity apprx} into Eq. \ref{Eq DR_GEP} we obtain the approximated expression for the dispersion relation:

\begin{equation}
\label{Eq DR_GEP cond}  
q_{GEP}=\frac{\varepsilon_0\varepsilon_{hBN_{eff}}\omega}{2\sigma_0f_m} \bigg(E-E_m+i\frac{\Gamma_m}{2}\bigg).
\end{equation}

Then the confinement factor and inverse damping ratio can be approximately expressed by (SI section E%appendix \ref{Section app GEP}
):

\begin{equation}
\label{Eq conf}
\lambda_0 / \lambda_{GEP}=\frac{\varepsilon_0\varepsilon_{hBN_{eff}}c}{2\sigma_0f_m} (E-E_m) ,
\end{equation}

\begin{equation}
\label{Eq FOM}        \gamma_p^{-1}=2\frac{E -E_m}{\Gamma_m}.
\end{equation}

In addition, by neglecting losses, the dispersion relation in Eq. \ref{Eq DR_GEP cond} can be expressed in terms of $\omega(q_{GEP})$:

\begin{equation}
\label{Eq DR_GEP in E terms}  
\omega=\frac{\omega_m}{2} + \sqrt{\frac{2\sigma_0f_m}{\varepsilon_0\varepsilon_{hBN_{eff}} \hbar} q_{GEP} + \bigg(\frac{\omega_m}{2} \bigg)^2} ,
\end{equation}

where $\omega_m=\frac{E_m}{\hbar}$ is the angular frequency of the exciton resonance.
Despite the fundamental plasmonic nature of the GEPs, as they are based on the existence of positive values of the imaginary part of the conductivity, Eq. \ref{Eq DR_GEP in E terms} differs significantly from the commonly known dispersion relation expected in all two-dimensional
electronic systems \cite{Stern1967PolarizabilityGas}. This behavior stems directly from the unique excitonic nature of these plasmon-like polaritons in comparison to plasmons in electronic systems. A clear shift in the exciton energy and momentum can be seen in the dispersion relation, introducing a new universal relation for surface polaritons in 2D excitonic systems.

We note that the polaritonic properties of the GEPs can be similarly evaluated for every bias voltage and every temperature for which the BBLG or BRTG presents positive values of the imaginary part of the conductivity. It can be seen from Fig. \ref{fig cond} (c) that the oscillator strength increases with bias voltage, suggesting that larger confinement factors can be obtained for lower voltages (Eq. \ref{Eq conf}). However, the effective permittivity of hBN, which also affects the confinement factor via Eq. \ref{Eq conf}, is frequency dependent %, as will be further discussed in section \ref{section hybridization}, 
and its real part obtains maximum values below and close to the hBN's Reststrahlen bands \cite{Dai2014TunableNitride}. By examining the dependence of Eq. \ref{Eq conf} on the voltage, we find that the largest confinement is obtained for a bias voltage of $V=52\,\mathrm{meV}$ for BBLG and of $V=55\,\mathrm{meV}$ for BRTG, which correspond to an exciton energy of $E_m \approx 90\,\mathrm{meV}$, just below the lower Reststrahlen band of hBN% (Fig. \ref{fig_hybrid_cond} (a))
. From Eq. \ref{Eq FOM}, we get that the inverse damping ratio increases with decreasing temperature due to a reduction in exciton-phonon interaction. Since $T=0\,\mathrm{K}$ is a theoretical limit we choose $T=4\,\mathrm{K}$ as the optimal temperature.

\section{Nonlocal Effects}
\label{section nonlocal}

\begin{figure}[t]
    \centering
    \hspace*{-13pt} % Shift the figure slightly to the left
    \begin{tikzpicture}
    \node[anchor=south west,inner sep=0] (image) 
    at (0,0) 
    {\includegraphics[width=1.03\columnwidth]
    {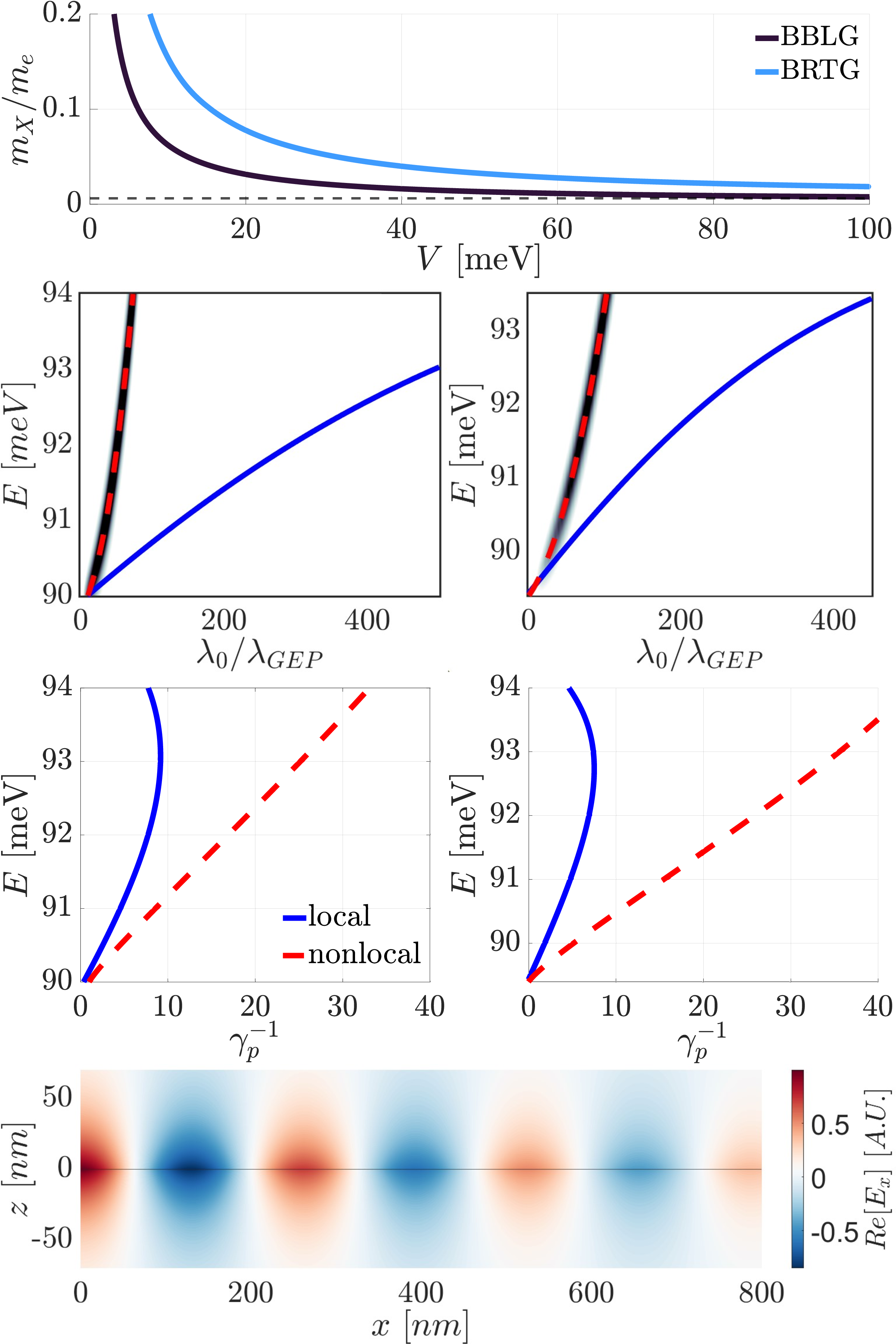}};
    %\label{''}
    \begin{scope}[x={(image.south east)},y={(image.north west)}]
    \node at (0,0.98) {(a)};
    \node at (0,0.77) {(b)};
    \node at (0.52,0.77) {(c)};
    \node at (0,0.48) {(d)};
    \node at (0.52,0.48) {(e)};
    \node at (0,0.2) {(f)};
    \end{scope}
    \end{tikzpicture}
    \caption{Nonlocal corrections. (a) Voltage dependence of the exciton's effective mass, calculated from Eq. \ref{Eq mass} for BBLG and numerically for BRTG, normalized by the electron mass. 
     Confinement factor of GEP $\lambda_0 / \lambda_{GEP}$, for BBLG for $V=52\,\mathrm{meV}$ (b) and for BRTG for $V=55\,\mathrm{meV}$ (c), both for $T=4\,\mathrm{K}$, including nonlocal effects, calculated from Eq. \ref{Eq NL DR} (dashed red line) and simulated using a TMM (colormap), compared with the local case (blue line) calculated from Eq. \ref{Eq DR_GEP}. 
    Inverse damping ratio of GEP $\gamma_p^{-1}=Re(q_{GEP}) / Im(q_{GEP})$, for BBLG (d) and BRTG (e).
    (f) The electromagnetic field distribution of GEP in BBLG, including nonlocal effects. The nonlocal corrections cause less confined modes with increased inverse damping ratio.}
    \label{ fig_nonlocal}
\end{figure}
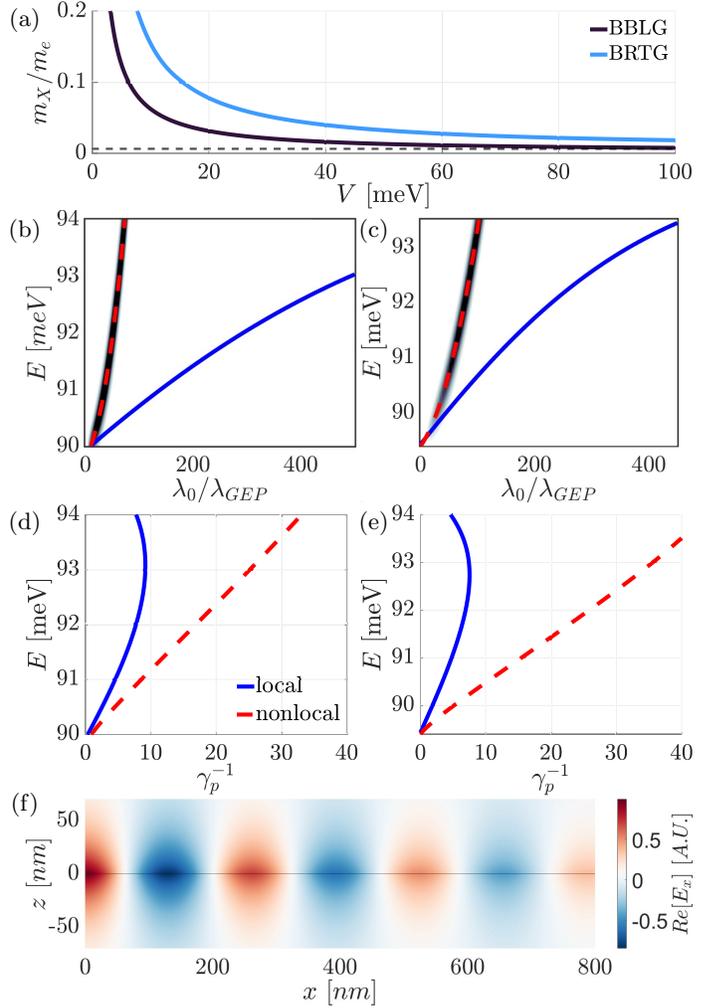

Next, we introduce the nonlocal correction to enhance the robustness and accuracy of our theoretical framework. Nonlocal effects describe the interaction between excitations in the material that cause the current at point $\boldsymbol{r}$ to depend on the value of the field at another point $\boldsymbol{r'} \neq \boldsymbol{r}$. In the momentum space, this leads to an optical response of the material that depends on the momentum as well as energy. In semiconductors, nonlocal effects are a result of coupling between neighboring excitons \cite{Eini2022Valley-polarizedFrequencies, Agarwal1974ElectromagneticMedia, Hopfield1963TheoreticalCrystals} and leads to a shift of the exciton energy described by the term $\frac{\hbar^2 q^2}{2 m_X}$, where $m_X$ is the exciton's effective mass. This approach holds for $\frac{\hbar^2 q^2}{2 m_X} \ll E_m$, such that the correction to the exciton energy is small.
To examine the conductivity near the main resonance where Eq. \ref{Eq conductivity apprx} holds, we neglect again the contribution of the additional resonances by assuming $|E-E_n| \gg \Gamma_n$ for $n \neq m$ and $|E-E_m|\lesssim\Gamma_m$. 
Thus, we obtain the expression for the nonlocal conductivity by setting $E_m \rightarrow E_m+\frac{\hbar^2 q^2}{2 m_X}$ in Eq. \ref{Eq conductivity apprx}:

\begin{equation} \label{Eq nonlocal conductivity}
    \sigma=\frac{4i\sigma_0 f_m}{E-(E_m+\frac{\hbar^2 q^2}{2 m_X})+i\frac{\Gamma_m}{2}} ,
\end{equation}

with $m_X$ in the BBLG given by (SI section F%appendix \ref{section app eff mass}
):

\begin{equation} \label{Eq mass}
    m_X=\frac{\gamma_1 (4V^2+\gamma_1^2)^\frac{3}{2}}{16V v_f ^2 (2V^2 +\gamma_1 ^2)},
\end{equation}

where $v_f=\frac{3a \gamma_0}{2 \hbar}$ is the Fermi velocity, $a=2.46 \,\text{\AA}$ is the lattice constant of graphene, $\gamma_0=3\,\mathrm{eV}$ and $\gamma_1=0.4\,\mathrm{eV}$ are the nearest neighbors intralayer and interlayer hoppings. 
The exciton's effective mass of BRTG is calculated in the same approach as in SI section F%appendix \ref{section app eff mass}
, but numerically in this case, owing to the complexity of the band structure.
The dependence of the effective mass on the bias voltage is presented in Fig. \ref{ fig_nonlocal} (a) and decreases with increasing voltage while $\frac{\hbar^2 q^2}{2 m_X}$ increases with increasing voltage, making the nonlocal correction to the conductivity in Eq. \ref{Eq nonlocal conductivity} more significant. 

Substituting the nonlocal conductivity of Eq. \ref{Eq nonlocal conductivity} into the dispersion relation in Eq. \ref{Eq DR_GEP} allows us to obtain the more accurate expression to the dispersion relation in the form of the quadratic equation: 

\begin{align}
    \label{Eq NL quad eq}
    \frac{\hbar^2 q^2}{2 m_X}+\frac{2\sigma_0 f_m}{\varepsilon_0 \varepsilon_{hBN_{eff}} \omega} q - \bigg(E-E_m+i\frac{\Gamma_m}{2}\bigg) = 0 , 
\end{align}

with the following physical solution of the nonlocal-corrected dispersion relation: 

\begin{align}
 \label{Eq NL DR} 
    \begin{split}
    q_{GEP} &=
    -\frac{2\sigma_0 f_m m_X}{\varepsilon_0 \varepsilon_{hBN_{eff}} \hbar^2 \omega}+    \\
    &\quad +\sqrt{\bigg(\frac{2\sigma_0 f_m m_X}{\varepsilon_0 \varepsilon_{hBN_{eff}} \hbar^2 \omega} \bigg)^2+ \frac{2 m_X}{\hbar ^2} \bigg( E-E_m+i \frac{\Gamma_m}{2} \bigg)}.
    \end{split}
\end{align}

Fig. \ref{ fig_nonlocal} (b) and (c) show the comparison between the confinement factors obtained via the local analytic solution (Eq. \ref{Eq DR_GEP}) and nonlocal analytic solution (Eq. \ref{Eq NL DR}) for BBLG and BRTG, respectively. It can be seen that the nonlocal dispersion relation yields GEP modes that are less confined by up to an order of magnitude compared to the local model, but that are still larger than the maximal limit of graphene plasmons \cite{Alonso-Gonzalez2016AcousticNanoscopy, Lundeberg2017TuningPlasmonics, Iranzo2018ProbingHeterostructure, Epstein2020Far-fieldVolumes, Menabde2021Real-spaceDeposition}. In addition, the inverse damping ratio, $\gamma_p^{-1}$, improves by a factor of four (Fig. \ref{ fig_nonlocal} (d) and (e)), reaching values that would enable their observation in cryo-SNOM experiments \cite{Ni2018FundamentalPlasmonics}.
We note that this behavior of a decreased confinement and increased inverse damping ratio due to nonlocal corrections is consistent with similar observations in other polaritonic systems \cite{Maier2007Plasmonics:Applications, Lundeberg2017TuningPlasmonics, Eini2022Valley-polarizedFrequencies, Gershuni2024In-planeLoss}. The resulting electromagnetic field distribution is presented in Fig. \ref{ fig_nonlocal} (f), and exhibits the expected decrease in confinement (larger polaritonic wavelength) and increase in the inverse damping ratio (larger propagation length), compared to Fig. \ref{fig_DR} (e). 

 In conclusion, we have predicted the existence of GEPs collective excitations in BBLG and BRTG at FIR frequencies, which follow a universal dispersion relation law for surface polaritons in 2D excitonic systems. We have studied their electrically tunable polaritonic properties
 taking into account the appropriate nonlocal corrections, and found that they exhibit promising polariton properties that would enable their observation in cryo-SNOM experiments. Such electrically tunable interband collective excitations in graphene systems may open up new research avenues for polaritonic phenomena that are based on excitonic response. 

\section*{Acknowledments}

I.E. acknowledges the support of the European Union (ERC, TOP-BLG, Project No. 101078192). N.M.R.P acknowledges support from the European Union through project EIC PATHFINDER OPEN project No. 101129661-ADAPTATION, and  the Portuguese Foundation for Science and Technology (FCT) in the framework of the Strategic Funding UIDB/04650/2020, COMPETE 2020, PORTUGAL 2020, FEDER, and through project PTDC/FIS-MAC/2045/2021.

\bibliography{GEPs.bib}

\begin{thebibliography}{66}%
\makeatletter
\providecommand \@ifxundefined [1]{%
 \@ifx{#1\undefined}
}%
\providecommand \@ifnum [1]{%
 \ifnum #1\expandafter \@firstoftwo
 \else \expandafter \@secondoftwo
 \fi
}%
\providecommand \@ifx [1]{%
 \ifx #1\expandafter \@firstoftwo
 \else \expandafter \@secondoftwo
 \fi
}%
\providecommand \natexlab [1]{#1}%
\providecommand \enquote  [1]{``#1''}%
\providecommand \bibnamefont  [1]{#1}%
\providecommand \bibfnamefont [1]{#1}%
\providecommand \citenamefont [1]{#1}%
\providecommand \href@noop [0]{\@secondoftwo}%
\providecommand \href [0]{\begingroup \@sanitize@url \@href}%
\providecommand \@href[1]{\@@startlink{#1}\@@href}%
\providecommand \@@href[1]{\endgroup#1\@@endlink}%
\providecommand \@sanitize@url [0]{\catcode `\\12\catcode `\$12\catcode `\&12\catcode `\#12\catcode `\^12\catcode `\_12\catcode `\%12\relax}%
\providecommand \@@startlink[1]{}%
\providecommand \@@endlink[0]{}%
\providecommand \url  [0]{\begingroup\@sanitize@url \@url }%
\providecommand \@url [1]{\endgroup\@href {#1}{\urlprefix }}%
\providecommand \urlprefix  [0]{URL }%
\providecommand \Eprint [0]{\href }%
\providecommand \doibase [0]{https://doi.org/}%
\providecommand \selectlanguage [0]{\@gobble}%
\providecommand \bibinfo  [0]{\@secondoftwo}%
\providecommand \bibfield  [0]{\@secondoftwo}%
\providecommand \translation [1]{[#1]}%
\providecommand \BibitemOpen [0]{}%
\providecommand \bibitemStop [0]{}%
\providecommand \bibitemNoStop [0]{.\EOS\space}%
\providecommand \EOS [0]{\spacefactor3000\relax}%
\providecommand \BibitemShut  [1]{\csname bibitem#1\endcsname}%
\let\auto@bib@innerbib\@empty
%</preamble>
\bibitem [{\citenamefont {Basov}\ \emph {et~al.}(2020)\citenamefont {Basov}, \citenamefont {Asenjo-Garcia}, \citenamefont {Schuck}, \citenamefont {Zhu},\ and\ \citenamefont {Rubio}}]{Basov2020PolaritonPanorama}%
  \BibitemOpen
  \bibfield  {author} {\bibinfo {author} {\bibfnamefont {D.~N.}\ \bibnamefont {Basov}}, \bibinfo {author} {\bibfnamefont {A.}~\bibnamefont {Asenjo-Garcia}}, \bibinfo {author} {\bibfnamefont {P.~J.}\ \bibnamefont {Schuck}}, \bibinfo {author} {\bibfnamefont {X.}~\bibnamefont {Zhu}},\ and\ \bibinfo {author} {\bibfnamefont {A.}~\bibnamefont {Rubio}},\ }\bibfield  {title} {\bibinfo {title} {{Polariton panorama}},\ }\href {https://doi.org/10.1515/NANOPH-2020-0449/ASSET/GRAPHIC/J{\_}NANOPH-2020-0449{\_}FIG{\_}002.JPG} {\bibfield  {journal} {\bibinfo  {journal} {Nanophotonics}\ }\textbf {\bibinfo {volume} {10}},\ \bibinfo {pages} {549} (\bibinfo {year} {2020})}\BibitemShut {NoStop}%
\bibitem [{\citenamefont {Basov}\ \emph {et~al.}(2016)\citenamefont {Basov}, \citenamefont {Fogler},\ and\ \citenamefont {Garc{\'{i}}a De~Abajo}}]{Basov2016PolaritonsMaterials}%
  \BibitemOpen
  \bibfield  {author} {\bibinfo {author} {\bibfnamefont {D.~N.}\ \bibnamefont {Basov}}, \bibinfo {author} {\bibfnamefont {M.~M.}\ \bibnamefont {Fogler}},\ and\ \bibinfo {author} {\bibfnamefont {F.~J.}\ \bibnamefont {Garc{\'{i}}a De~Abajo}},\ }\bibfield  {title} {\bibinfo {title} {{Polaritons in van der Waals materials}},\ }\bibfield  {journal} {\bibinfo  {journal} {Science}\ }\textbf {\bibinfo {volume} {354}},\ \href {https://doi.org/10.1126/SCIENCE.AAG1992} {10.1126/SCIENCE.AAG1992} (\bibinfo {year} {2016})\BibitemShut {NoStop}%
\bibitem [{\citenamefont {Low}\ \emph {et~al.}(2016)\citenamefont {Low}, \citenamefont {Chaves}, \citenamefont {Caldwell}, \citenamefont {Kumar}, \citenamefont {Fang}, \citenamefont {Avouris}, \citenamefont {Heinz}, \citenamefont {Guinea}, \citenamefont {Martin-Moreno},\ and\ \citenamefont {Koppens}}]{Low2016PolaritonsMaterials}%
  \BibitemOpen
  \bibfield  {author} {\bibinfo {author} {\bibfnamefont {T.}~\bibnamefont {Low}}, \bibinfo {author} {\bibfnamefont {A.}~\bibnamefont {Chaves}}, \bibinfo {author} {\bibfnamefont {J.~D.}\ \bibnamefont {Caldwell}}, \bibinfo {author} {\bibfnamefont {A.}~\bibnamefont {Kumar}}, \bibinfo {author} {\bibfnamefont {N.~X.}\ \bibnamefont {Fang}}, \bibinfo {author} {\bibfnamefont {P.}~\bibnamefont {Avouris}}, \bibinfo {author} {\bibfnamefont {T.~F.}\ \bibnamefont {Heinz}}, \bibinfo {author} {\bibfnamefont {F.}~\bibnamefont {Guinea}}, \bibinfo {author} {\bibfnamefont {L.}~\bibnamefont {Martin-Moreno}},\ and\ \bibinfo {author} {\bibfnamefont {F.}~\bibnamefont {Koppens}},\ }\bibfield  {title} {\bibinfo {title} {{Polaritons in layered two-dimensional materials}},\ }\href {https://doi.org/10.1038/nmat4792} {\bibfield  {journal} {\bibinfo  {journal} {Nature Materials 2016 16:2}\ }\textbf {\bibinfo {volume} {16}},\ \bibinfo {pages} {182} (\bibinfo {year} {2016})}\BibitemShut {NoStop}%
\bibitem [{\citenamefont {Jablan}\ \emph {et~al.}(2009)\citenamefont {Jablan}, \citenamefont {Buljan},\ and\ \citenamefont {Solja{\v{c}}i{\'{c}}}}]{Jablan2009PlasmonicsFrequencies}%
  \BibitemOpen
  \bibfield  {author} {\bibinfo {author} {\bibfnamefont {M.}~\bibnamefont {Jablan}}, \bibinfo {author} {\bibfnamefont {H.}~\bibnamefont {Buljan}},\ and\ \bibinfo {author} {\bibfnamefont {M.}~\bibnamefont {Solja{\v{c}}i{\'{c}}}},\ }\bibfield  {title} {\bibinfo {title} {{Plasmonics in graphene at infrared frequencies}},\ }\href {https://doi.org/10.1103/PHYSREVB.80.245435/FIGURES/5/MEDIUM} {\bibfield  {journal} {\bibinfo  {journal} {Physical Review B - Condensed Matter and Materials Physics}\ }\textbf {\bibinfo {volume} {80}},\ \bibinfo {pages} {245435} (\bibinfo {year} {2009})}\BibitemShut {NoStop}%
\bibitem [{\citenamefont {Koppens}\ \emph {et~al.}(2011)\citenamefont {Koppens}, \citenamefont {Chang},\ and\ \citenamefont {Garc{\'{i}}a De~Abajo}}]{Koppens2011}%
  \BibitemOpen
  \bibfield  {author} {\bibinfo {author} {\bibfnamefont {F.~H.}\ \bibnamefont {Koppens}}, \bibinfo {author} {\bibfnamefont {D.~E.}\ \bibnamefont {Chang}},\ and\ \bibinfo {author} {\bibfnamefont {F.~J.}\ \bibnamefont {Garc{\'{i}}a De~Abajo}},\ }\bibfield  {title} {\bibinfo {title} {{Graphene plasmonics: A platform for strong light-matter interactions}},\ }\href {https://doi.org/10.1021/NL201771H/SUPPL{\_}FILE/NL201771H{\_}SI{\_}001.PDF} {\bibfield  {journal} {\bibinfo  {journal} {Nano Letters}\ }\textbf {\bibinfo {volume} {11}},\ \bibinfo {pages} {3370} (\bibinfo {year} {2011})}\BibitemShut {NoStop}%
\bibitem [{\citenamefont {Dai}\ \emph {et~al.}(2014)\citenamefont {Dai}, \citenamefont {Fei}, \citenamefont {Ma}, \citenamefont {Rodin}, \citenamefont {Wagner}, \citenamefont {McLeod}, \citenamefont {Liu}, \citenamefont {Gannett}, \citenamefont {Regan}, \citenamefont {Watanabe}, \citenamefont {Taniguchi}, \citenamefont {Thiemens}, \citenamefont {Dominguez}, \citenamefont {Castro~Neto}, \citenamefont {Zettl}, \citenamefont {Keilmann}, \citenamefont {Jarillo-Herrero}, \citenamefont {Fogler},\ and\ \citenamefont {Basov}}]{Dai2014TunableNitride}%
  \BibitemOpen
  \bibfield  {author} {\bibinfo {author} {\bibfnamefont {S.}~\bibnamefont {Dai}}, \bibinfo {author} {\bibfnamefont {Z.}~\bibnamefont {Fei}}, \bibinfo {author} {\bibfnamefont {Q.}~\bibnamefont {Ma}}, \bibinfo {author} {\bibfnamefont {A.~S.}\ \bibnamefont {Rodin}}, \bibinfo {author} {\bibfnamefont {M.}~\bibnamefont {Wagner}}, \bibinfo {author} {\bibfnamefont {A.~S.}\ \bibnamefont {McLeod}}, \bibinfo {author} {\bibfnamefont {M.~K.}\ \bibnamefont {Liu}}, \bibinfo {author} {\bibfnamefont {W.}~\bibnamefont {Gannett}}, \bibinfo {author} {\bibfnamefont {W.}~\bibnamefont {Regan}}, \bibinfo {author} {\bibfnamefont {K.}~\bibnamefont {Watanabe}}, \bibinfo {author} {\bibfnamefont {T.}~\bibnamefont {Taniguchi}}, \bibinfo {author} {\bibfnamefont {M.}~\bibnamefont {Thiemens}}, \bibinfo {author} {\bibfnamefont {G.}~\bibnamefont {Dominguez}}, \bibinfo {author} {\bibfnamefont {A.~H.}\ \bibnamefont {Castro~Neto}}, \bibinfo {author} {\bibfnamefont {A.}~\bibnamefont {Zettl}}, \bibinfo {author} {\bibfnamefont {F.}~\bibnamefont
  {Keilmann}}, \bibinfo {author} {\bibfnamefont {P.}~\bibnamefont {Jarillo-Herrero}}, \bibinfo {author} {\bibfnamefont {M.~M.}\ \bibnamefont {Fogler}},\ and\ \bibinfo {author} {\bibfnamefont {D.~N.}\ \bibnamefont {Basov}},\ }\bibfield  {title} {\bibinfo {title} {{Tunable phonon polaritons in atomically thin van der Waals crystals of boron nitride}},\ }\href {https://doi.org/10.1126/SCIENCE.1246833} {\bibfield  {journal} {\bibinfo  {journal} {Science}\ }\textbf {\bibinfo {volume} {343}},\ \bibinfo {pages} {1125} (\bibinfo {year} {2014})}\BibitemShut {NoStop}%
\bibitem [{\citenamefont {Caldwell}\ \emph {et~al.}(2015)\citenamefont {Caldwell}, \citenamefont {Lindsay}, \citenamefont {Giannini}, \citenamefont {Vurgaftman}, \citenamefont {Reinecke}, \citenamefont {Maier},\ and\ \citenamefont {Glembocki}}]{Caldwell2015Low-lossPolaritons}%
  \BibitemOpen
  \bibfield  {author} {\bibinfo {author} {\bibfnamefont {J.~D.}\ \bibnamefont {Caldwell}}, \bibinfo {author} {\bibfnamefont {L.}~\bibnamefont {Lindsay}}, \bibinfo {author} {\bibfnamefont {V.}~\bibnamefont {Giannini}}, \bibinfo {author} {\bibfnamefont {I.}~\bibnamefont {Vurgaftman}}, \bibinfo {author} {\bibfnamefont {T.~L.}\ \bibnamefont {Reinecke}}, \bibinfo {author} {\bibfnamefont {S.~A.}\ \bibnamefont {Maier}},\ and\ \bibinfo {author} {\bibfnamefont {O.~J.}\ \bibnamefont {Glembocki}},\ }\bibfield  {title} {\bibinfo {title} {{Low-loss, infrared and terahertz nanophotonics using surface phonon polaritons}},\ }\href {https://doi.org/10.1515/NANOPH-2014-0003} {\bibfield  {journal} {\bibinfo  {journal} {Nanophotonics}\ }\textbf {\bibinfo {volume} {4}},\ \bibinfo {pages} {44} (\bibinfo {year} {2015})}\BibitemShut {NoStop}%
\bibitem [{\citenamefont {Caldwell}\ \emph {et~al.}(2014)\citenamefont {Caldwell}, \citenamefont {Kretinin}, \citenamefont {Chen}, \citenamefont {Giannini}, \citenamefont {Fogler}, \citenamefont {Francescato}, \citenamefont {Ellis}, \citenamefont {Tischler}, \citenamefont {Woods}, \citenamefont {Giles}, \citenamefont {Hong}, \citenamefont {Watanabe}, \citenamefont {Taniguchi}, \citenamefont {Maier},\ and\ \citenamefont {Novoselov}}]{Caldwell2014Sub-diffractionalNitride}%
  \BibitemOpen
  \bibfield  {author} {\bibinfo {author} {\bibfnamefont {J.~D.}\ \bibnamefont {Caldwell}}, \bibinfo {author} {\bibfnamefont {A.~V.}\ \bibnamefont {Kretinin}}, \bibinfo {author} {\bibfnamefont {Y.}~\bibnamefont {Chen}}, \bibinfo {author} {\bibfnamefont {V.}~\bibnamefont {Giannini}}, \bibinfo {author} {\bibfnamefont {M.~M.}\ \bibnamefont {Fogler}}, \bibinfo {author} {\bibfnamefont {Y.}~\bibnamefont {Francescato}}, \bibinfo {author} {\bibfnamefont {C.~T.}\ \bibnamefont {Ellis}}, \bibinfo {author} {\bibfnamefont {J.~G.}\ \bibnamefont {Tischler}}, \bibinfo {author} {\bibfnamefont {C.~R.}\ \bibnamefont {Woods}}, \bibinfo {author} {\bibfnamefont {A.~J.}\ \bibnamefont {Giles}}, \bibinfo {author} {\bibfnamefont {M.}~\bibnamefont {Hong}}, \bibinfo {author} {\bibfnamefont {K.}~\bibnamefont {Watanabe}}, \bibinfo {author} {\bibfnamefont {T.}~\bibnamefont {Taniguchi}}, \bibinfo {author} {\bibfnamefont {S.~A.}\ \bibnamefont {Maier}},\ and\ \bibinfo {author} {\bibfnamefont {K.~S.}\ \bibnamefont {Novoselov}},\ }\bibfield
  {title} {\bibinfo {title} {{Sub-diffractional volume-confined polaritons in the natural hyperbolic material hexagonal boron nitride}},\ }\href {https://doi.org/10.1038/ncomms6221} {\bibfield  {journal} {\bibinfo  {journal} {Nature Communications 2014 5:1}\ }\textbf {\bibinfo {volume} {5}},\ \bibinfo {pages} {1} (\bibinfo {year} {2014})}\BibitemShut {NoStop}%
\bibitem [{\citenamefont {Schneider}\ \emph {et~al.}(2018)\citenamefont {Schneider}, \citenamefont {Glazov}, \citenamefont {Korn}, \citenamefont {H{\"{o}}fling},\ and\ \citenamefont {Urbaszek}}]{Schneider2018Two-dimensionalCoupling}%
  \BibitemOpen
  \bibfield  {author} {\bibinfo {author} {\bibfnamefont {C.}~\bibnamefont {Schneider}}, \bibinfo {author} {\bibfnamefont {M.~M.}\ \bibnamefont {Glazov}}, \bibinfo {author} {\bibfnamefont {T.}~\bibnamefont {Korn}}, \bibinfo {author} {\bibfnamefont {S.}~\bibnamefont {H{\"{o}}fling}},\ and\ \bibinfo {author} {\bibfnamefont {B.}~\bibnamefont {Urbaszek}},\ }\bibfield  {title} {\bibinfo {title} {{Two-dimensional semiconductors in the regime of strong light-matter coupling}},\ }\href {https://doi.org/10.1038/s41467-018-04866-6} {\bibfield  {journal} {\bibinfo  {journal} {Nature Communications 2018 9:1}\ }\textbf {\bibinfo {volume} {9}},\ \bibinfo {pages} {1} (\bibinfo {year} {2018})}\BibitemShut {NoStop}%
\bibitem [{\citenamefont {Maier}(2007)}]{Maier2007Plasmonics:Applications}%
  \BibitemOpen
  \bibfield  {author} {\bibinfo {author} {\bibfnamefont {S.~A.}\ \bibnamefont {Maier}},\ }\bibfield  {title} {\bibinfo {title} {{Plasmonics: Fundamentals and applications}},\ }\href {https://doi.org/10.1007/0-387-37825-1/COVER} {\bibfield  {journal} {\bibinfo  {journal} {Plasmonics: Fundamentals and Applications}\ ,\ \bibinfo {pages} {1}} (\bibinfo {year} {2007})}\BibitemShut {NoStop}%
\bibitem [{\citenamefont {Barnes}\ \emph {et~al.}(2003)\citenamefont {Barnes}, \citenamefont {Dereux},\ and\ \citenamefont {Ebbesen}}]{Barnes2003SurfaceOptics}%
  \BibitemOpen
  \bibfield  {author} {\bibinfo {author} {\bibfnamefont {W.~L.}\ \bibnamefont {Barnes}}, \bibinfo {author} {\bibfnamefont {A.}~\bibnamefont {Dereux}},\ and\ \bibinfo {author} {\bibfnamefont {T.~W.}\ \bibnamefont {Ebbesen}},\ }\bibfield  {title} {\bibinfo {title} {{Surface plasmon subwavelength optics}},\ }\href {https://doi.org/10.1038/nature01937} {\bibfield  {journal} {\bibinfo  {journal} {Nature 2003 424:6950}\ }\textbf {\bibinfo {volume} {424}},\ \bibinfo {pages} {824} (\bibinfo {year} {2003})}\BibitemShut {NoStop}%
\bibitem [{\citenamefont {Ebbesen}\ \emph {et~al.}(1998)\citenamefont {Ebbesen}, \citenamefont {Lezec}, \citenamefont {Ghaemi}, \citenamefont {Thio},\ and\ \citenamefont {Wolff}}]{Ebbesen1998ExtraordinaryArrays}%
  \BibitemOpen
  \bibfield  {author} {\bibinfo {author} {\bibfnamefont {T.~W.}\ \bibnamefont {Ebbesen}}, \bibinfo {author} {\bibfnamefont {H.~J.}\ \bibnamefont {Lezec}}, \bibinfo {author} {\bibfnamefont {H.~F.}\ \bibnamefont {Ghaemi}}, \bibinfo {author} {\bibfnamefont {T.}~\bibnamefont {Thio}},\ and\ \bibinfo {author} {\bibfnamefont {P.~A.}\ \bibnamefont {Wolff}},\ }\bibfield  {title} {\bibinfo {title} {{Extraordinary optical transmission through sub-wavelength hole arrays}},\ }\href {https://doi.org/10.1038/35570} {\bibfield  {journal} {\bibinfo  {journal} {Nature 1998 391:6668}\ }\textbf {\bibinfo {volume} {391}},\ \bibinfo {pages} {667} (\bibinfo {year} {1998})}\BibitemShut {NoStop}%
\bibitem [{\citenamefont {Epstein}\ \emph {et~al.}(2016)\citenamefont {Epstein}, \citenamefont {Tsur},\ and\ \citenamefont {Arie}}]{Epstein2016Surface-plasmonHolography}%
  \BibitemOpen
  \bibfield  {author} {\bibinfo {author} {\bibfnamefont {I.}~\bibnamefont {Epstein}}, \bibinfo {author} {\bibfnamefont {Y.}~\bibnamefont {Tsur}},\ and\ \bibinfo {author} {\bibfnamefont {A.}~\bibnamefont {Arie}},\ }\bibfield  {title} {\bibinfo {title} {{Surface-plasmon wavefront and spectral shaping by near-field holography}},\ }\href {https://doi.org/10.1002/LPOR.201500242} {\bibfield  {journal} {\bibinfo  {journal} {Laser {\&} Photonics Reviews}\ }\textbf {\bibinfo {volume} {10}},\ \bibinfo {pages} {360} (\bibinfo {year} {2016})}\BibitemShut {NoStop}%
\bibitem [{\citenamefont {Agranovich}(2014)}]{Agranovich2014SurfacePolaritons}%
  \BibitemOpen
  \bibfield  {author} {\bibinfo {author} {\bibfnamefont {V.~M.}\ \bibnamefont {Agranovich}},\ }\href {https://books.google.com/books/about/Surface_Polaritons.html?id=wKJpV8c2ZtQC} {\emph {\bibinfo {title} {{Surface Polaritons}}}}\ (\bibinfo  {publisher} {Elsevier Science},\ \bibinfo {year} {2014})\ p.\ \bibinfo {pages} {734}\BibitemShut {NoStop}%
\bibitem [{\citenamefont {Lagois}\ and\ \citenamefont {Fischer}(1976)}]{Lagois1976ExperimentalPolaritons}%
  \BibitemOpen
  \bibfield  {author} {\bibinfo {author} {\bibfnamefont {J.}~\bibnamefont {Lagois}}\ and\ \bibinfo {author} {\bibfnamefont {B.}~\bibnamefont {Fischer}},\ }\bibfield  {title} {\bibinfo {title} {{Experimental Observation of Surface Exciton Polaritons}},\ }\href {https://doi.org/10.1103/PhysRevLett.36.680} {\bibfield  {journal} {\bibinfo  {journal} {Physical Review Letters}\ }\textbf {\bibinfo {volume} {36}},\ \bibinfo {pages} {680} (\bibinfo {year} {1976})}\BibitemShut {NoStop}%
\bibitem [{\citenamefont {Lagois}\ and\ \citenamefont {Fischer}(1978)}]{Lagois1978DispersionPolaritons}%
  \BibitemOpen
  \bibfield  {author} {\bibinfo {author} {\bibfnamefont {J.}~\bibnamefont {Lagois}}\ and\ \bibinfo {author} {\bibfnamefont {B.}~\bibnamefont {Fischer}},\ }\bibfield  {title} {\bibinfo {title} {{Dispersion theory of surface-exciton polaritons}},\ }\href {https://doi.org/10.1103/PhysRevB.17.3814} {\bibfield  {journal} {\bibinfo  {journal} {Physical Review B}\ }\textbf {\bibinfo {volume} {17}},\ \bibinfo {pages} {3814} (\bibinfo {year} {1978})}\BibitemShut {NoStop}%
\bibitem [{\citenamefont {Elrafei}\ \emph {et~al.}(2024)\citenamefont {Elrafei}, \citenamefont {Raziman}, \citenamefont {De~Vega}, \citenamefont {Garc{\'{i}}a De~Abajo},\ and\ \citenamefont {Curto}}]{Elrafei2024GuidingSuperlattices}%
  \BibitemOpen
  \bibfield  {author} {\bibinfo {author} {\bibfnamefont {S.~A.}\ \bibnamefont {Elrafei}}, \bibinfo {author} {\bibfnamefont {T.~V.}\ \bibnamefont {Raziman}}, \bibinfo {author} {\bibfnamefont {S.}~\bibnamefont {De~Vega}}, \bibinfo {author} {\bibfnamefont {F.~J.}\ \bibnamefont {Garc{\'{i}}a De~Abajo}},\ and\ \bibinfo {author} {\bibfnamefont {A.~G.}\ \bibnamefont {Curto}},\ }\bibfield  {title} {\bibinfo {title} {{Guiding light with surface exciton-polaritons in atomically thin superlattices}},\ }\href {https://doi.org/10.1515/NANOPH-2024-0075/ASSET/GRAPHIC/J{\_}NANOPH-2024-0075{\_}FIG{\_}007.JPG} {\bibfield  {journal} {\bibinfo  {journal} {Nanophotonics}\ }\textbf {\bibinfo {volume} {13}},\ \bibinfo {pages} {3101} (\bibinfo {year} {2024})}\BibitemShut {NoStop}%
\bibitem [{\citenamefont {Tokura}\ and\ \citenamefont {Koda}(1982)}]{Tokura1982SurfaceZnO}%
  \BibitemOpen
  \bibfield  {author} {\bibinfo {author} {\bibfnamefont {Y.}~\bibnamefont {Tokura}}\ and\ \bibinfo {author} {\bibfnamefont {T.}~\bibnamefont {Koda}},\ }\bibfield  {title} {\bibinfo {title} {{Surface Exciton Polariton in ZnO}},\ }\href {https://doi.org/10.1143/JPSJ.51.2934} {\bibfield  {journal} {\bibinfo  {journal} {https://doi.org/10.1143/JPSJ.51.2934}\ }\textbf {\bibinfo {volume} {51}},\ \bibinfo {pages} {2934} (\bibinfo {year} {1982})}\BibitemShut {NoStop}%
\bibitem [{\citenamefont {Chen}\ \emph {et~al.}(2012)\citenamefont {Chen}, \citenamefont {Badioli}, \citenamefont {Alonso-Gonz{\'{a}}lez}, \citenamefont {Thongrattanasiri}, \citenamefont {Huth}, \citenamefont {Osmond}, \citenamefont {Spasenovi{\'{c}}}, \citenamefont {Centeno}, \citenamefont {Pesquera}, \citenamefont {Godignon}, \citenamefont {Zurutuza~Elorza}, \citenamefont {Camara}, \citenamefont {Garc{\'{i}}a}, \citenamefont {Hillenbrand},\ and\ \citenamefont {Koppens}}]{Chen2012OpticalPlasmons}%
  \BibitemOpen
  \bibfield  {author} {\bibinfo {author} {\bibfnamefont {J.}~\bibnamefont {Chen}}, \bibinfo {author} {\bibfnamefont {M.}~\bibnamefont {Badioli}}, \bibinfo {author} {\bibfnamefont {P.}~\bibnamefont {Alonso-Gonz{\'{a}}lez}}, \bibinfo {author} {\bibfnamefont {S.}~\bibnamefont {Thongrattanasiri}}, \bibinfo {author} {\bibfnamefont {F.}~\bibnamefont {Huth}}, \bibinfo {author} {\bibfnamefont {J.}~\bibnamefont {Osmond}}, \bibinfo {author} {\bibfnamefont {M.}~\bibnamefont {Spasenovi{\'{c}}}}, \bibinfo {author} {\bibfnamefont {A.}~\bibnamefont {Centeno}}, \bibinfo {author} {\bibfnamefont {A.}~\bibnamefont {Pesquera}}, \bibinfo {author} {\bibfnamefont {P.}~\bibnamefont {Godignon}}, \bibinfo {author} {\bibfnamefont {A.}~\bibnamefont {Zurutuza~Elorza}}, \bibinfo {author} {\bibfnamefont {N.}~\bibnamefont {Camara}}, \bibinfo {author} {\bibfnamefont {F.~J.}\ \bibnamefont {Garc{\'{i}}a}}, \bibinfo {author} {\bibfnamefont {R.}~\bibnamefont {Hillenbrand}},\ and\ \bibinfo {author} {\bibfnamefont {F.~H.}\ \bibnamefont
  {Koppens}},\ }\bibfield  {title} {\bibinfo {title} {{Optical nano-imaging of gate-tunable graphene plasmons}},\ }\href {https://doi.org/10.1038/nature11254} {\bibfield  {journal} {\bibinfo  {journal} {Nature 2012 487:7405}\ }\textbf {\bibinfo {volume} {487}},\ \bibinfo {pages} {77} (\bibinfo {year} {2012})}\BibitemShut {NoStop}%
\bibitem [{\citenamefont {Fei}\ \emph {et~al.}(2012)\citenamefont {Fei}, \citenamefont {Rodin}, \citenamefont {Andreev}, \citenamefont {Bao}, \citenamefont {McLeod}, \citenamefont {Wagner}, \citenamefont {Zhang}, \citenamefont {Zhao}, \citenamefont {Thiemens}, \citenamefont {Dominguez}, \citenamefont {Fogler}, \citenamefont {Castro~Neto}, \citenamefont {Lau}, \citenamefont {Keilmann},\ and\ \citenamefont {Basov}}]{Fei2012Gate-tuningNano-imaging}%
  \BibitemOpen
  \bibfield  {author} {\bibinfo {author} {\bibfnamefont {Z.}~\bibnamefont {Fei}}, \bibinfo {author} {\bibfnamefont {A.~S.}\ \bibnamefont {Rodin}}, \bibinfo {author} {\bibfnamefont {G.~O.}\ \bibnamefont {Andreev}}, \bibinfo {author} {\bibfnamefont {W.}~\bibnamefont {Bao}}, \bibinfo {author} {\bibfnamefont {A.~S.}\ \bibnamefont {McLeod}}, \bibinfo {author} {\bibfnamefont {M.}~\bibnamefont {Wagner}}, \bibinfo {author} {\bibfnamefont {L.~M.}\ \bibnamefont {Zhang}}, \bibinfo {author} {\bibfnamefont {Z.}~\bibnamefont {Zhao}}, \bibinfo {author} {\bibfnamefont {M.}~\bibnamefont {Thiemens}}, \bibinfo {author} {\bibfnamefont {G.}~\bibnamefont {Dominguez}}, \bibinfo {author} {\bibfnamefont {M.~M.}\ \bibnamefont {Fogler}}, \bibinfo {author} {\bibfnamefont {A.~H.}\ \bibnamefont {Castro~Neto}}, \bibinfo {author} {\bibfnamefont {C.~N.}\ \bibnamefont {Lau}}, \bibinfo {author} {\bibfnamefont {F.}~\bibnamefont {Keilmann}},\ and\ \bibinfo {author} {\bibfnamefont {D.~N.}\ \bibnamefont {Basov}},\ }\bibfield  {title} {\bibinfo
  {title} {{Gate-tuning of graphene plasmons revealed by infrared nano-imaging}},\ }\href {https://doi.org/10.1038/nature11253} {\bibfield  {journal} {\bibinfo  {journal} {Nature 2012 487:7405}\ }\textbf {\bibinfo {volume} {487}},\ \bibinfo {pages} {82} (\bibinfo {year} {2012})}\BibitemShut {NoStop}%
\bibitem [{\citenamefont {Hesp}\ \emph {et~al.}(2021)\citenamefont {Hesp}, \citenamefont {Torre}, \citenamefont {Rodan-Legrain}, \citenamefont {Novelli}, \citenamefont {Cao}, \citenamefont {Carr}, \citenamefont {Fang}, \citenamefont {Stepanov}, \citenamefont {Barcons-Ruiz}, \citenamefont {Herzig~Sheinfux}, \citenamefont {Watanabe}, \citenamefont {Taniguchi}, \citenamefont {Efetov}, \citenamefont {Kaxiras}, \citenamefont {Jarillo-Herrero}, \citenamefont {Polini},\ and\ \citenamefont {Koppens}}]{Hesp2021ObservationGraphene}%
  \BibitemOpen
  \bibfield  {author} {\bibinfo {author} {\bibfnamefont {N.~C.}\ \bibnamefont {Hesp}}, \bibinfo {author} {\bibfnamefont {I.}~\bibnamefont {Torre}}, \bibinfo {author} {\bibfnamefont {D.}~\bibnamefont {Rodan-Legrain}}, \bibinfo {author} {\bibfnamefont {P.}~\bibnamefont {Novelli}}, \bibinfo {author} {\bibfnamefont {Y.}~\bibnamefont {Cao}}, \bibinfo {author} {\bibfnamefont {S.}~\bibnamefont {Carr}}, \bibinfo {author} {\bibfnamefont {S.}~\bibnamefont {Fang}}, \bibinfo {author} {\bibfnamefont {P.}~\bibnamefont {Stepanov}}, \bibinfo {author} {\bibfnamefont {D.}~\bibnamefont {Barcons-Ruiz}}, \bibinfo {author} {\bibfnamefont {H.}~\bibnamefont {Herzig~Sheinfux}}, \bibinfo {author} {\bibfnamefont {K.}~\bibnamefont {Watanabe}}, \bibinfo {author} {\bibfnamefont {T.}~\bibnamefont {Taniguchi}}, \bibinfo {author} {\bibfnamefont {D.~K.}\ \bibnamefont {Efetov}}, \bibinfo {author} {\bibfnamefont {E.}~\bibnamefont {Kaxiras}}, \bibinfo {author} {\bibfnamefont {P.}~\bibnamefont {Jarillo-Herrero}}, \bibinfo {author} {\bibfnamefont
  {M.}~\bibnamefont {Polini}},\ and\ \bibinfo {author} {\bibfnamefont {F.~H.}\ \bibnamefont {Koppens}},\ }\bibfield  {title} {\bibinfo {title} {{Observation of interband collective excitations in twisted bilayer graphene}},\ }\href {https://doi.org/10.1038/s41567-021-01327-8} {\bibfield  {journal} {\bibinfo  {journal} {Nature Physics 2021 17:10}\ }\textbf {\bibinfo {volume} {17}},\ \bibinfo {pages} {1162} (\bibinfo {year} {2021})}\BibitemShut {NoStop}%
\bibitem [{\citenamefont {Hillenbrand}\ \emph {et~al.}(2002)\citenamefont {Hillenbrand}, \citenamefont {Taubner},\ and\ \citenamefont {Keilmann}}]{Hillenbrand2002Phonon-enhancedScale}%
  \BibitemOpen
  \bibfield  {author} {\bibinfo {author} {\bibfnamefont {R.}~\bibnamefont {Hillenbrand}}, \bibinfo {author} {\bibfnamefont {T.}~\bibnamefont {Taubner}},\ and\ \bibinfo {author} {\bibfnamefont {F.}~\bibnamefont {Keilmann}},\ }\bibfield  {title} {\bibinfo {title} {{Phonon-enhanced light matter interaction at the nanometre scale}},\ }\href {https://doi.org/10.1038/NATURE00899} {\bibfield  {journal} {\bibinfo  {journal} {Nature}\ }\textbf {\bibinfo {volume} {418}},\ \bibinfo {pages} {159} (\bibinfo {year} {2002})}\BibitemShut {NoStop}%
\bibitem [{\citenamefont {Caldwell}\ \emph {et~al.}(2013)\citenamefont {Caldwell}, \citenamefont {Glembocki}, \citenamefont {Francescato}, \citenamefont {Sharac}, \citenamefont {Giannini}, \citenamefont {Bezares}, \citenamefont {Long}, \citenamefont {Owrutsky}, \citenamefont {Vurgaftman}, \citenamefont {Tischler}, \citenamefont {Wheeler}, \citenamefont {Bassim}, \citenamefont {Shirey}, \citenamefont {Kasica},\ and\ \citenamefont {Maier}}]{Caldwell2013Low-lossResonators}%
  \BibitemOpen
  \bibfield  {author} {\bibinfo {author} {\bibfnamefont {J.~D.}\ \bibnamefont {Caldwell}}, \bibinfo {author} {\bibfnamefont {O.~J.}\ \bibnamefont {Glembocki}}, \bibinfo {author} {\bibfnamefont {Y.}~\bibnamefont {Francescato}}, \bibinfo {author} {\bibfnamefont {N.}~\bibnamefont {Sharac}}, \bibinfo {author} {\bibfnamefont {V.}~\bibnamefont {Giannini}}, \bibinfo {author} {\bibfnamefont {F.~J.}\ \bibnamefont {Bezares}}, \bibinfo {author} {\bibfnamefont {J.~P.}\ \bibnamefont {Long}}, \bibinfo {author} {\bibfnamefont {J.~C.}\ \bibnamefont {Owrutsky}}, \bibinfo {author} {\bibfnamefont {I.}~\bibnamefont {Vurgaftman}}, \bibinfo {author} {\bibfnamefont {J.~G.}\ \bibnamefont {Tischler}}, \bibinfo {author} {\bibfnamefont {V.~D.}\ \bibnamefont {Wheeler}}, \bibinfo {author} {\bibfnamefont {N.~D.}\ \bibnamefont {Bassim}}, \bibinfo {author} {\bibfnamefont {L.~M.}\ \bibnamefont {Shirey}}, \bibinfo {author} {\bibfnamefont {R.}~\bibnamefont {Kasica}},\ and\ \bibinfo {author} {\bibfnamefont {S.~A.}\ \bibnamefont {Maier}},\
  }\bibfield  {title} {\bibinfo {title} {{Low-loss, extreme subdiffraction photon confinement via silicon carbide localized surface phonon polariton resonators}},\ }\href {https://doi.org/10.1021/NL401590G} {\bibfield  {journal} {\bibinfo  {journal} {Nano letters}\ }\textbf {\bibinfo {volume} {13}},\ \bibinfo {pages} {3690} (\bibinfo {year} {2013})}\BibitemShut {NoStop}%
\bibitem [{\citenamefont {Weisbuch}\ \emph {et~al.}(1992)\citenamefont {Weisbuch}, \citenamefont {Nishioka}, \citenamefont {Ishikawa},\ and\ \citenamefont {Arakawa}}]{Weisbuch1992ObservationMicrocavity}%
  \BibitemOpen
  \bibfield  {author} {\bibinfo {author} {\bibfnamefont {C.}~\bibnamefont {Weisbuch}}, \bibinfo {author} {\bibfnamefont {M.}~\bibnamefont {Nishioka}}, \bibinfo {author} {\bibfnamefont {A.}~\bibnamefont {Ishikawa}},\ and\ \bibinfo {author} {\bibfnamefont {Y.}~\bibnamefont {Arakawa}},\ }\bibfield  {title} {\bibinfo {title} {{Observation of the coupled exciton-photon mode splitting in a semiconductor quantum microcavity}},\ }\href {https://doi.org/10.1103/PhysRevLett.69.3314} {\bibfield  {journal} {\bibinfo  {journal} {Physical Review Letters}\ }\textbf {\bibinfo {volume} {69}},\ \bibinfo {pages} {3314} (\bibinfo {year} {1992})}\BibitemShut {NoStop}%
\bibitem [{\citenamefont {Kavokin}(2010)}]{Kavokin2010Exciton-polaritonsPerspectives}%
  \BibitemOpen
  \bibfield  {author} {\bibinfo {author} {\bibfnamefont {A.}~\bibnamefont {Kavokin}},\ }\bibfield  {title} {\bibinfo {title} {{Exciton-polaritons in microcavities: Recent discoveries and perspectives}},\ }\href {https://doi.org/10.1002/PSSB.200983955} {\bibfield  {journal} {\bibinfo  {journal} {physica status solidi (b)}\ }\textbf {\bibinfo {volume} {247}},\ \bibinfo {pages} {1898} (\bibinfo {year} {2010})}\BibitemShut {NoStop}%
\bibitem [{\citenamefont {Deng}\ \emph {et~al.}(2010)\citenamefont {Deng}, \citenamefont {Haug},\ and\ \citenamefont {Yamamoto}}]{Deng2010Exciton-polaritonCondensation}%
  \BibitemOpen
  \bibfield  {author} {\bibinfo {author} {\bibfnamefont {H.}~\bibnamefont {Deng}}, \bibinfo {author} {\bibfnamefont {H.}~\bibnamefont {Haug}},\ and\ \bibinfo {author} {\bibfnamefont {Y.}~\bibnamefont {Yamamoto}},\ }\bibfield  {title} {\bibinfo {title} {{Exciton-polariton Bose-Einstein condensation}},\ }\href {https://doi.org/10.1103/REVMODPHYS.82.1489/FIGURES/49/MEDIUM} {\bibfield  {journal} {\bibinfo  {journal} {Reviews of Modern Physics}\ }\textbf {\bibinfo {volume} {82}},\ \bibinfo {pages} {1489} (\bibinfo {year} {2010})}\BibitemShut {NoStop}%
\bibitem [{\citenamefont {Timofeev}\ and\ \citenamefont {Sanvitto}(2012)}]{Timofeev2012ExcitonFrontiers}%
  \BibitemOpen
  \bibfield  {author} {\bibinfo {author} {\bibfnamefont {V.}~\bibnamefont {Timofeev}}\ and\ \bibinfo {author} {\bibfnamefont {D.}~\bibnamefont {Sanvitto}},\ }\href {https://books.google.com/books/about/Exciton_Polaritons_in_Microcavities.html?hl=iw&id=3ybfbv8d3NUC} {\emph {\bibinfo {title} {{Exciton polaritons in microcavities : new frontiers}}}}\ (\bibinfo  {publisher} {Springer},\ \bibinfo {year} {2012})\ p.\ \bibinfo {pages} {401}\BibitemShut {NoStop}%
\bibitem [{\citenamefont {Yamamoto}\ and\ \citenamefont {Imamoglu}(1999)}]{Yamamoto1999MesoscopicOptics}%
  \BibitemOpen
  \bibfield  {author} {\bibinfo {author} {\bibfnamefont {Y.}~\bibnamefont {Yamamoto}}\ and\ \bibinfo {author} {\bibfnamefont {A.}~\bibnamefont {Imamoglu}},\ }\href {https://www.wiley.com/en-us/Mesoscopic+Quantum+Optics-p-9780471148746} {\emph {\bibinfo {title} {{Mesoscopic Quantum Optics}}}}\ (\bibinfo  {publisher} {John Wiley},\ \bibinfo {year} {1999})\ p.\ \bibinfo {pages} {320}\BibitemShut {NoStop}%
\bibitem [{\citenamefont {Liu}\ \emph {et~al.}(2014)\citenamefont {Liu}, \citenamefont {Galfsky}, \citenamefont {Sun}, \citenamefont {Xia}, \citenamefont {Lin}, \citenamefont {Lee}, \citenamefont {K{\'{e}}na-Cohen},\ and\ \citenamefont {Menon}}]{Liu2014StrongCrystals}%
  \BibitemOpen
  \bibfield  {author} {\bibinfo {author} {\bibfnamefont {X.}~\bibnamefont {Liu}}, \bibinfo {author} {\bibfnamefont {T.}~\bibnamefont {Galfsky}}, \bibinfo {author} {\bibfnamefont {Z.}~\bibnamefont {Sun}}, \bibinfo {author} {\bibfnamefont {F.}~\bibnamefont {Xia}}, \bibinfo {author} {\bibfnamefont {E.-c.}\ \bibnamefont {Lin}}, \bibinfo {author} {\bibfnamefont {Y.-H.}\ \bibnamefont {Lee}}, \bibinfo {author} {\bibfnamefont {S.}~\bibnamefont {K{\'{e}}na-Cohen}},\ and\ \bibinfo {author} {\bibfnamefont {V.~M.}\ \bibnamefont {Menon}},\ }\bibfield  {title} {\bibinfo {title} {{Strong light-matter coupling in two-dimensional atomic crystals}},\ }\href {https://doi.org/10.1038/nphoton.2014.304} {\bibfield  {journal} {\bibinfo  {journal} {Nature Photonics}\ }\textbf {\bibinfo {volume} {9}},\ \bibinfo {pages} {30} (\bibinfo {year} {2014})}\BibitemShut {NoStop}%
\bibitem [{\citenamefont {Alonso-Gonz{\'{a}}lez}\ \emph {et~al.}(2016)\citenamefont {Alonso-Gonz{\'{a}}lez}, \citenamefont {Nikitin}, \citenamefont {Gao}, \citenamefont {Woessner}, \citenamefont {Lundeberg}, \citenamefont {Principi}, \citenamefont {Forcellini}, \citenamefont {Yan}, \citenamefont {V{\'{e}}lez}, \citenamefont {Huber}, \citenamefont {Watanabe}, \citenamefont {Taniguchi}, \citenamefont {Casanova}, \citenamefont {Hueso}, \citenamefont {Polini}, \citenamefont {Hone}, \citenamefont {Koppens},\ and\ \citenamefont {Hillenbrand}}]{Alonso-Gonzalez2016AcousticNanoscopy}%
  \BibitemOpen
  \bibfield  {author} {\bibinfo {author} {\bibfnamefont {P.}~\bibnamefont {Alonso-Gonz{\'{a}}lez}}, \bibinfo {author} {\bibfnamefont {A.~Y.}\ \bibnamefont {Nikitin}}, \bibinfo {author} {\bibfnamefont {Y.}~\bibnamefont {Gao}}, \bibinfo {author} {\bibfnamefont {A.}~\bibnamefont {Woessner}}, \bibinfo {author} {\bibfnamefont {M.~B.}\ \bibnamefont {Lundeberg}}, \bibinfo {author} {\bibfnamefont {A.}~\bibnamefont {Principi}}, \bibinfo {author} {\bibfnamefont {N.}~\bibnamefont {Forcellini}}, \bibinfo {author} {\bibfnamefont {W.}~\bibnamefont {Yan}}, \bibinfo {author} {\bibfnamefont {S.}~\bibnamefont {V{\'{e}}lez}}, \bibinfo {author} {\bibfnamefont {A.~J.}\ \bibnamefont {Huber}}, \bibinfo {author} {\bibfnamefont {K.}~\bibnamefont {Watanabe}}, \bibinfo {author} {\bibfnamefont {T.}~\bibnamefont {Taniguchi}}, \bibinfo {author} {\bibfnamefont {F.}~\bibnamefont {Casanova}}, \bibinfo {author} {\bibfnamefont {L.~E.}\ \bibnamefont {Hueso}}, \bibinfo {author} {\bibfnamefont {M.}~\bibnamefont {Polini}}, \bibinfo {author}
  {\bibfnamefont {J.}~\bibnamefont {Hone}}, \bibinfo {author} {\bibfnamefont {F.~H.~L.}\ \bibnamefont {Koppens}},\ and\ \bibinfo {author} {\bibfnamefont {R.}~\bibnamefont {Hillenbrand}},\ }\bibfield  {title} {\bibinfo {title} {{Acoustic terahertz graphene plasmons revealed by photocurrent nanoscopy}},\ }\href {https://doi.org/10.1038/nnano.2016.185} {\bibfield  {journal} {\bibinfo  {journal} {Nature Nanotechnology 2016 12:1}\ }\textbf {\bibinfo {volume} {12}},\ \bibinfo {pages} {31} (\bibinfo {year} {2016})}\BibitemShut {NoStop}%
\bibitem [{\citenamefont {Alonso-Gonz{\'{a}}lez}\ \emph {et~al.}(2014)\citenamefont {Alonso-Gonz{\'{a}}lez}, \citenamefont {Nikitin}, \citenamefont {Golmar}, \citenamefont {Centeno}, \citenamefont {Pesquera}, \citenamefont {V{\'{e}}lez}, \citenamefont {Chen}, \citenamefont {Navickaite}, \citenamefont {Koppens}, \citenamefont {Zurutuza}, \citenamefont {Casanova}, \citenamefont {Hueso},\ and\ \citenamefont {Hillenbrand}}]{Alonso-Gonzalez2014ControllingPatterns}%
  \BibitemOpen
  \bibfield  {author} {\bibinfo {author} {\bibfnamefont {P.}~\bibnamefont {Alonso-Gonz{\'{a}}lez}}, \bibinfo {author} {\bibfnamefont {A.~Y.}\ \bibnamefont {Nikitin}}, \bibinfo {author} {\bibfnamefont {F.}~\bibnamefont {Golmar}}, \bibinfo {author} {\bibfnamefont {A.}~\bibnamefont {Centeno}}, \bibinfo {author} {\bibfnamefont {A.}~\bibnamefont {Pesquera}}, \bibinfo {author} {\bibfnamefont {S.}~\bibnamefont {V{\'{e}}lez}}, \bibinfo {author} {\bibfnamefont {J.}~\bibnamefont {Chen}}, \bibinfo {author} {\bibfnamefont {G.}~\bibnamefont {Navickaite}}, \bibinfo {author} {\bibfnamefont {F.}~\bibnamefont {Koppens}}, \bibinfo {author} {\bibfnamefont {A.}~\bibnamefont {Zurutuza}}, \bibinfo {author} {\bibfnamefont {F.}~\bibnamefont {Casanova}}, \bibinfo {author} {\bibfnamefont {L.~E.}\ \bibnamefont {Hueso}},\ and\ \bibinfo {author} {\bibfnamefont {R.}~\bibnamefont {Hillenbrand}},\ }\bibfield  {title} {\bibinfo {title} {{Controlling graphene plasmons with resonant metal antennas and spatial conductivity patterns}},\ }\href
  {https://doi.org/10.1126/SCIENCE.1253202/SUPPL{\_}FILE/ALONSO.GONZALEZ.SM.PDF} {\bibfield  {journal} {\bibinfo  {journal} {Science}\ }\textbf {\bibinfo {volume} {344}},\ \bibinfo {pages} {1369} (\bibinfo {year} {2014})}\BibitemShut {NoStop}%
\bibitem [{\citenamefont {Ju}\ \emph {et~al.}(2011)\citenamefont {Ju}, \citenamefont {Geng}, \citenamefont {Horng}, \citenamefont {Girit}, \citenamefont {Martin}, \citenamefont {Hao}, \citenamefont {Bechtel}, \citenamefont {Liang}, \citenamefont {Zettl}, \citenamefont {Shen},\ and\ \citenamefont {Wang}}]{Ju2011GrapheneMetamaterials}%
  \BibitemOpen
  \bibfield  {author} {\bibinfo {author} {\bibfnamefont {L.}~\bibnamefont {Ju}}, \bibinfo {author} {\bibfnamefont {B.}~\bibnamefont {Geng}}, \bibinfo {author} {\bibfnamefont {J.}~\bibnamefont {Horng}}, \bibinfo {author} {\bibfnamefont {C.}~\bibnamefont {Girit}}, \bibinfo {author} {\bibfnamefont {M.}~\bibnamefont {Martin}}, \bibinfo {author} {\bibfnamefont {Z.}~\bibnamefont {Hao}}, \bibinfo {author} {\bibfnamefont {H.~A.}\ \bibnamefont {Bechtel}}, \bibinfo {author} {\bibfnamefont {X.}~\bibnamefont {Liang}}, \bibinfo {author} {\bibfnamefont {A.}~\bibnamefont {Zettl}}, \bibinfo {author} {\bibfnamefont {Y.~R.}\ \bibnamefont {Shen}},\ and\ \bibinfo {author} {\bibfnamefont {F.}~\bibnamefont {Wang}},\ }\bibfield  {title} {\bibinfo {title} {{Graphene plasmonics for tunable terahertz metamaterials}},\ }\href {https://doi.org/10.1038/nnano.2011.146} {\bibfield  {journal} {\bibinfo  {journal} {Nature Nanotechnology 2011 6:10}\ }\textbf {\bibinfo {volume} {6}},\ \bibinfo {pages} {630} (\bibinfo {year}
  {2011})}\BibitemShut {NoStop}%
\bibitem [{\citenamefont {Ni}\ \emph {et~al.}(2018)\citenamefont {Ni}, \citenamefont {McLeod}, \citenamefont {Sun}, \citenamefont {Wang}, \citenamefont {Xiong}, \citenamefont {Post}, \citenamefont {Sunku}, \citenamefont {Jiang}, \citenamefont {Hone}, \citenamefont {Dean}, \citenamefont {Fogler},\ and\ \citenamefont {Basov}}]{Ni2018FundamentalPlasmonics}%
  \BibitemOpen
  \bibfield  {author} {\bibinfo {author} {\bibfnamefont {G.~X.}\ \bibnamefont {Ni}}, \bibinfo {author} {\bibfnamefont {A.~S.}\ \bibnamefont {McLeod}}, \bibinfo {author} {\bibfnamefont {Z.}~\bibnamefont {Sun}}, \bibinfo {author} {\bibfnamefont {L.}~\bibnamefont {Wang}}, \bibinfo {author} {\bibfnamefont {L.}~\bibnamefont {Xiong}}, \bibinfo {author} {\bibfnamefont {K.~W.}\ \bibnamefont {Post}}, \bibinfo {author} {\bibfnamefont {S.~S.}\ \bibnamefont {Sunku}}, \bibinfo {author} {\bibfnamefont {B.~Y.}\ \bibnamefont {Jiang}}, \bibinfo {author} {\bibfnamefont {J.}~\bibnamefont {Hone}}, \bibinfo {author} {\bibfnamefont {C.~R.}\ \bibnamefont {Dean}}, \bibinfo {author} {\bibfnamefont {M.~M.}\ \bibnamefont {Fogler}},\ and\ \bibinfo {author} {\bibfnamefont {D.~N.}\ \bibnamefont {Basov}},\ }\bibfield  {title} {\bibinfo {title} {{Fundamental limits to graphene plasmonics}},\ }\href {https://doi.org/10.1038/s41586-018-0136-9} {\bibfield  {journal} {\bibinfo  {journal} {Nature 2018 557:7706}\ }\textbf {\bibinfo {volume}
  {557}},\ \bibinfo {pages} {530} (\bibinfo {year} {2018})}\BibitemShut {NoStop}%
\bibitem [{\citenamefont {Sun}\ \emph {et~al.}(2017)\citenamefont {Sun}, \citenamefont {Gu}, \citenamefont {Ghazaryan}, \citenamefont {Shotan}, \citenamefont {Considine}, \citenamefont {Dollar}, \citenamefont {Chakraborty}, \citenamefont {Liu}, \citenamefont {Ghaemi}, \citenamefont {K{\'{e}}na-Cohen},\ and\ \citenamefont {Menon}}]{Sun2017OpticalPolaritons}%
  \BibitemOpen
  \bibfield  {author} {\bibinfo {author} {\bibfnamefont {Z.}~\bibnamefont {Sun}}, \bibinfo {author} {\bibfnamefont {J.}~\bibnamefont {Gu}}, \bibinfo {author} {\bibfnamefont {A.}~\bibnamefont {Ghazaryan}}, \bibinfo {author} {\bibfnamefont {Z.}~\bibnamefont {Shotan}}, \bibinfo {author} {\bibfnamefont {C.~R.}\ \bibnamefont {Considine}}, \bibinfo {author} {\bibfnamefont {M.}~\bibnamefont {Dollar}}, \bibinfo {author} {\bibfnamefont {B.}~\bibnamefont {Chakraborty}}, \bibinfo {author} {\bibfnamefont {X.}~\bibnamefont {Liu}}, \bibinfo {author} {\bibfnamefont {P.}~\bibnamefont {Ghaemi}}, \bibinfo {author} {\bibfnamefont {S.}~\bibnamefont {K{\'{e}}na-Cohen}},\ and\ \bibinfo {author} {\bibfnamefont {V.~M.}\ \bibnamefont {Menon}},\ }\bibfield  {title} {\bibinfo {title} {{Optical control of room-temperature valley polaritons}},\ }\href {https://doi.org/10.1038/nphoton.2017.121} {\bibfield  {journal} {\bibinfo  {journal} {Nature Photonics 2017 11:8}\ }\textbf {\bibinfo {volume} {11}},\ \bibinfo {pages} {491} (\bibinfo
  {year} {2017})}\BibitemShut {NoStop}%
\bibitem [{\citenamefont {Epstein}\ \emph {et~al.}(2020{\natexlab{a}})\citenamefont {Epstein}, \citenamefont {Terr{\'{e}}s}, \citenamefont {Chaves}, \citenamefont {Pusapati}, \citenamefont {Rhodes}, \citenamefont {Frank}, \citenamefont {Zimmermann}, \citenamefont {Qin}, \citenamefont {Watanabe}, \citenamefont {Taniguchi}, \citenamefont {Giessen}, \citenamefont {Tongay}, \citenamefont {Hone}, \citenamefont {Peres},\ and\ \citenamefont {Koppens}}]{Epstein2020Near-UnityCavityb}%
  \BibitemOpen
  \bibfield  {author} {\bibinfo {author} {\bibfnamefont {I.}~\bibnamefont {Epstein}}, \bibinfo {author} {\bibfnamefont {B.}~\bibnamefont {Terr{\'{e}}s}}, \bibinfo {author} {\bibfnamefont {A.~J.}\ \bibnamefont {Chaves}}, \bibinfo {author} {\bibfnamefont {V.~V.}\ \bibnamefont {Pusapati}}, \bibinfo {author} {\bibfnamefont {D.~A.}\ \bibnamefont {Rhodes}}, \bibinfo {author} {\bibfnamefont {B.}~\bibnamefont {Frank}}, \bibinfo {author} {\bibfnamefont {V.}~\bibnamefont {Zimmermann}}, \bibinfo {author} {\bibfnamefont {Y.}~\bibnamefont {Qin}}, \bibinfo {author} {\bibfnamefont {K.}~\bibnamefont {Watanabe}}, \bibinfo {author} {\bibfnamefont {T.}~\bibnamefont {Taniguchi}}, \bibinfo {author} {\bibfnamefont {H.}~\bibnamefont {Giessen}}, \bibinfo {author} {\bibfnamefont {S.}~\bibnamefont {Tongay}}, \bibinfo {author} {\bibfnamefont {J.~C.}\ \bibnamefont {Hone}}, \bibinfo {author} {\bibfnamefont {N.~M.}\ \bibnamefont {Peres}},\ and\ \bibinfo {author} {\bibfnamefont {F.~H.}\ \bibnamefont {Koppens}},\ }\bibfield  {title}
  {\bibinfo {title} {{Near-unity light absorption in a monolayer ws2 van der waals heterostructure cavity}},\ }\href {https://doi.org/10.1021/ACS.NANOLETT.0C00492/ASSET/IMAGES/LARGE/NL0C00492{\_}0004.JPEG} {\bibfield  {journal} {\bibinfo  {journal} {Nano Letters}\ }\textbf {\bibinfo {volume} {20}},\ \bibinfo {pages} {3545} (\bibinfo {year} {2020}{\natexlab{a}})}\BibitemShut {NoStop}%
\bibitem [{\citenamefont {Horng}\ \emph {et~al.}(2020)\citenamefont {Horng}, \citenamefont {Martin}, \citenamefont {Chou}, \citenamefont {Courtade}, \citenamefont {Chang}, \citenamefont {Hsu}, \citenamefont {Wentzel}, \citenamefont {Ruth}, \citenamefont {Lu}, \citenamefont {Cundiff}, \citenamefont {Wang},\ and\ \citenamefont {Deng}}]{Horng2020PerfectCrystal}%
  \BibitemOpen
  \bibfield  {author} {\bibinfo {author} {\bibfnamefont {J.}~\bibnamefont {Horng}}, \bibinfo {author} {\bibfnamefont {E.~W.}\ \bibnamefont {Martin}}, \bibinfo {author} {\bibfnamefont {Y.~H.}\ \bibnamefont {Chou}}, \bibinfo {author} {\bibfnamefont {E.}~\bibnamefont {Courtade}}, \bibinfo {author} {\bibfnamefont {T.~C.}\ \bibnamefont {Chang}}, \bibinfo {author} {\bibfnamefont {C.~Y.}\ \bibnamefont {Hsu}}, \bibinfo {author} {\bibfnamefont {M.~H.}\ \bibnamefont {Wentzel}}, \bibinfo {author} {\bibfnamefont {H.~G.}\ \bibnamefont {Ruth}}, \bibinfo {author} {\bibfnamefont {T.~C.}\ \bibnamefont {Lu}}, \bibinfo {author} {\bibfnamefont {S.~T.}\ \bibnamefont {Cundiff}}, \bibinfo {author} {\bibfnamefont {F.}~\bibnamefont {Wang}},\ and\ \bibinfo {author} {\bibfnamefont {H.}~\bibnamefont {Deng}},\ }\bibfield  {title} {\bibinfo {title} {{Perfect absorption by an atomically thin crystal}},\ }\href {https://doi.org/10.1103/PHYSREVAPPLIED.14.024009/FIGURES/4/MEDIUM} {\bibfield  {journal} {\bibinfo  {journal} {Physical Review
  Applied}\ }\textbf {\bibinfo {volume} {14}},\ \bibinfo {pages} {024009} (\bibinfo {year} {2020})}\BibitemShut {NoStop}%
\bibitem [{\citenamefont {Fang}\ \emph {et~al.}(2019)\citenamefont {Fang}, \citenamefont {Han}, \citenamefont {Robert}, \citenamefont {Semina}, \citenamefont {Lagarde}, \citenamefont {Courtade}, \citenamefont {Taniguchi}, \citenamefont {Watanabe}, \citenamefont {Amand}, \citenamefont {Urbaszek}, \citenamefont {Glazov},\ and\ \citenamefont {Marie}}]{Fang2019ControlHeterostructures}%
  \BibitemOpen
  \bibfield  {author} {\bibinfo {author} {\bibfnamefont {H.~H.}\ \bibnamefont {Fang}}, \bibinfo {author} {\bibfnamefont {B.}~\bibnamefont {Han}}, \bibinfo {author} {\bibfnamefont {C.}~\bibnamefont {Robert}}, \bibinfo {author} {\bibfnamefont {M.~A.}\ \bibnamefont {Semina}}, \bibinfo {author} {\bibfnamefont {D.}~\bibnamefont {Lagarde}}, \bibinfo {author} {\bibfnamefont {E.}~\bibnamefont {Courtade}}, \bibinfo {author} {\bibfnamefont {T.}~\bibnamefont {Taniguchi}}, \bibinfo {author} {\bibfnamefont {K.}~\bibnamefont {Watanabe}}, \bibinfo {author} {\bibfnamefont {T.}~\bibnamefont {Amand}}, \bibinfo {author} {\bibfnamefont {B.}~\bibnamefont {Urbaszek}}, \bibinfo {author} {\bibfnamefont {M.~M.}\ \bibnamefont {Glazov}},\ and\ \bibinfo {author} {\bibfnamefont {X.}~\bibnamefont {Marie}},\ }\bibfield  {title} {\bibinfo {title} {{Control of the Exciton Radiative Lifetime in van der Waals Heterostructures}},\ }\href {https://doi.org/10.1103/PHYSREVLETT.123.067401/FIGURES/3/MEDIUM} {\bibfield  {journal} {\bibinfo  {journal}
  {Physical Review Letters}\ }\textbf {\bibinfo {volume} {123}},\ \bibinfo {pages} {067401} (\bibinfo {year} {2019})}\BibitemShut {NoStop}%
\bibitem [{\citenamefont {Horng}\ \emph {et~al.}(2019)\citenamefont {Horng}, \citenamefont {Chou}, \citenamefont {Chou}, \citenamefont {Chang}, \citenamefont {Hsu}, \citenamefont {Lu}, \citenamefont {Deng},\ and\ \citenamefont {Deng}}]{Horng2019EngineeringSemiconductors}%
  \BibitemOpen
  \bibfield  {author} {\bibinfo {author} {\bibfnamefont {J.}~\bibnamefont {Horng}}, \bibinfo {author} {\bibfnamefont {Y.-H.}\ \bibnamefont {Chou}}, \bibinfo {author} {\bibfnamefont {Y.-H.}\ \bibnamefont {Chou}}, \bibinfo {author} {\bibfnamefont {T.-C.}\ \bibnamefont {Chang}}, \bibinfo {author} {\bibfnamefont {C.-Y.}\ \bibnamefont {Hsu}}, \bibinfo {author} {\bibfnamefont {T.-C.}\ \bibnamefont {Lu}}, \bibinfo {author} {\bibfnamefont {H.}~\bibnamefont {Deng}},\ and\ \bibinfo {author} {\bibfnamefont {H.}~\bibnamefont {Deng}},\ }\bibfield  {title} {\bibinfo {title} {{Engineering radiative coupling of excitons in 2D semiconductors}},\ }\href {https://doi.org/10.1364/OPTICA.6.001443} {\bibfield  {journal} {\bibinfo  {journal} {Optica, Vol. 6, Issue 11, pp. 1443-1448}\ }\textbf {\bibinfo {volume} {6}},\ \bibinfo {pages} {1443} (\bibinfo {year} {2019})}\BibitemShut {NoStop}%
\bibitem [{\citenamefont {Henriques}\ \emph {et~al.}(2022)\citenamefont {Henriques}, \citenamefont {Epstein},\ and\ \citenamefont {Peres}}]{Henriques2022AbsorptionGraphene}%
  \BibitemOpen
  \bibfield  {author} {\bibinfo {author} {\bibfnamefont {J.~C.}\ \bibnamefont {Henriques}}, \bibinfo {author} {\bibfnamefont {I.}~\bibnamefont {Epstein}},\ and\ \bibinfo {author} {\bibfnamefont {N.~M.}\ \bibnamefont {Peres}},\ }\bibfield  {title} {\bibinfo {title} {{Absorption and optical selection rules of tunable excitons in biased bilayer graphene}},\ }\href {https://doi.org/10.1103/PHYSREVB.105.045411/FIGURES/6/MEDIUM} {\bibfield  {journal} {\bibinfo  {journal} {Physical Review B}\ }\textbf {\bibinfo {volume} {105}},\ \bibinfo {pages} {045411} (\bibinfo {year} {2022})}\BibitemShut {NoStop}%
\bibitem [{\citenamefont {Quintela}\ and\ \citenamefont {Peres}(2022)}]{Quintela2022TunableGraphene}%
  \BibitemOpen
  \bibfield  {author} {\bibinfo {author} {\bibfnamefont {M.~F.}\ \bibnamefont {Quintela}}\ and\ \bibinfo {author} {\bibfnamefont {N.~M.}\ \bibnamefont {Peres}},\ }\bibfield  {title} {\bibinfo {title} {{Tunable excitons in rhombohedral trilayer graphene}},\ }\href {https://doi.org/10.1103/PHYSREVB.105.205417/FIGURES/6/MEDIUM} {\bibfield  {journal} {\bibinfo  {journal} {Physical Review B}\ }\textbf {\bibinfo {volume} {105}},\ \bibinfo {pages} {205417} (\bibinfo {year} {2022})}\BibitemShut {NoStop}%
\bibitem [{\citenamefont {Ju}\ \emph {et~al.}(2017)\citenamefont {Ju}, \citenamefont {Wang}, \citenamefont {Cao}, \citenamefont {Taniguchi}, \citenamefont {Watanabe}, \citenamefont {Louie}, \citenamefont {Rana}, \citenamefont {Park}, \citenamefont {Hone}, \citenamefont {Wang},\ and\ \citenamefont {McEuen}}]{Ju2017TunableGraphene}%
  \BibitemOpen
  \bibfield  {author} {\bibinfo {author} {\bibfnamefont {L.}~\bibnamefont {Ju}}, \bibinfo {author} {\bibfnamefont {L.}~\bibnamefont {Wang}}, \bibinfo {author} {\bibfnamefont {T.}~\bibnamefont {Cao}}, \bibinfo {author} {\bibfnamefont {T.}~\bibnamefont {Taniguchi}}, \bibinfo {author} {\bibfnamefont {K.}~\bibnamefont {Watanabe}}, \bibinfo {author} {\bibfnamefont {S.~G.}\ \bibnamefont {Louie}}, \bibinfo {author} {\bibfnamefont {F.}~\bibnamefont {Rana}}, \bibinfo {author} {\bibfnamefont {J.}~\bibnamefont {Park}}, \bibinfo {author} {\bibfnamefont {J.}~\bibnamefont {Hone}}, \bibinfo {author} {\bibfnamefont {F.}~\bibnamefont {Wang}},\ and\ \bibinfo {author} {\bibfnamefont {P.~L.}\ \bibnamefont {McEuen}},\ }\bibfield  {title} {\bibinfo {title} {{Tunable excitons in bilayer graphene}},\ }\href {https://doi.org/10.1126/SCIENCE.AAM9175/SUPPL{\_}FILE/AAM9175{\_}JU{\_}SM.PDF} {\bibfield  {journal} {\bibinfo  {journal} {Science}\ }\textbf {\bibinfo {volume} {358}},\ \bibinfo {pages} {907} (\bibinfo {year}
  {2017})}\BibitemShut {NoStop}%
\bibitem [{\citenamefont {Park}\ and\ \citenamefont {Louie}(2010)}]{Park2010TunableGraphene}%
  \BibitemOpen
  \bibfield  {author} {\bibinfo {author} {\bibfnamefont {C.~H.}\ \bibnamefont {Park}}\ and\ \bibinfo {author} {\bibfnamefont {S.~G.}\ \bibnamefont {Louie}},\ }\bibfield  {title} {\bibinfo {title} {{Tunable excitons in biased bilayer graphene}},\ }\href {https://doi.org/10.1021/NL902932K/ASSET/IMAGES/MEDIUM/NL-2009-02932K{\_}0006.GIF} {\bibfield  {journal} {\bibinfo  {journal} {Nano Letters}\ }\textbf {\bibinfo {volume} {10}},\ \bibinfo {pages} {426} (\bibinfo {year} {2010})}\BibitemShut {NoStop}%
\bibitem [{\citenamefont {Ju}\ \emph {et~al.}(2020)\citenamefont {Ju}, \citenamefont {Wang}, \citenamefont {Li}, \citenamefont {Moon}, \citenamefont {Ozerov}, \citenamefont {Lu}, \citenamefont {Taniguchi}, \citenamefont {Watanabe}, \citenamefont {Mueller}, \citenamefont {Zhang}, \citenamefont {Smirnov}, \citenamefont {Rana},\ and\ \citenamefont {McEuen}}]{Ju2020UnconventionalGraphene}%
  \BibitemOpen
  \bibfield  {author} {\bibinfo {author} {\bibfnamefont {L.}~\bibnamefont {Ju}}, \bibinfo {author} {\bibfnamefont {L.}~\bibnamefont {Wang}}, \bibinfo {author} {\bibfnamefont {X.}~\bibnamefont {Li}}, \bibinfo {author} {\bibfnamefont {S.}~\bibnamefont {Moon}}, \bibinfo {author} {\bibfnamefont {M.}~\bibnamefont {Ozerov}}, \bibinfo {author} {\bibfnamefont {Z.}~\bibnamefont {Lu}}, \bibinfo {author} {\bibfnamefont {T.}~\bibnamefont {Taniguchi}}, \bibinfo {author} {\bibfnamefont {K.}~\bibnamefont {Watanabe}}, \bibinfo {author} {\bibfnamefont {E.}~\bibnamefont {Mueller}}, \bibinfo {author} {\bibfnamefont {F.}~\bibnamefont {Zhang}}, \bibinfo {author} {\bibfnamefont {D.}~\bibnamefont {Smirnov}}, \bibinfo {author} {\bibfnamefont {F.}~\bibnamefont {Rana}},\ and\ \bibinfo {author} {\bibfnamefont {P.~L.}\ \bibnamefont {McEuen}},\ }\bibfield  {title} {\bibinfo {title} {{Unconventional valley-dependent optical selection rules and landau level mixing in bilayer graphene}},\ }\href {https://doi.org/10.1038/s41467-020-16844-y}
  {\bibfield  {journal} {\bibinfo  {journal} {Nature Communications 2020 11:1}\ }\textbf {\bibinfo {volume} {11}},\ \bibinfo {pages} {1} (\bibinfo {year} {2020})}\BibitemShut {NoStop}%
\bibitem [{\citenamefont {Duarte}\ \emph {et~al.}(2024)\citenamefont {Duarte}, \citenamefont {Da~Costa}, \citenamefont {Peres}, \citenamefont {Teles},\ and\ \citenamefont {Chaves}}]{Duarte2024MoirePressure}%
  \BibitemOpen
  \bibfield  {author} {\bibinfo {author} {\bibfnamefont {V.~G.}\ \bibnamefont {Duarte}}, \bibinfo {author} {\bibfnamefont {D.~R.}\ \bibnamefont {Da~Costa}}, \bibinfo {author} {\bibfnamefont {N.~M.}\ \bibnamefont {Peres}}, \bibinfo {author} {\bibfnamefont {L.~K.}\ \bibnamefont {Teles}},\ and\ \bibinfo {author} {\bibfnamefont {A.~J.}\ \bibnamefont {Chaves}},\ }\bibfield  {title} {\bibinfo {title} {{Moir{\'{e}} excitons in biased twisted bilayer graphene under pressure}},\ }\href {https://doi.org/10.1103/PHYSREVB.110.035405/FIGURES/9/MEDIUM} {\bibfield  {journal} {\bibinfo  {journal} {Physical Review B}\ }\textbf {\bibinfo {volume} {110}},\ \bibinfo {pages} {035405} (\bibinfo {year} {2024})}\BibitemShut {NoStop}%
\bibitem [{\citenamefont {Castro}\ \emph {et~al.}(2007)\citenamefont {Castro}, \citenamefont {Novoselov}, \citenamefont {Morozov}, \citenamefont {Peres}, \citenamefont {Dos~Santos}, \citenamefont {Nilsson}, \citenamefont {Guinea}, \citenamefont {Geim},\ and\ \citenamefont {Neto}}]{Castro2007BiasedEffect}%
  \BibitemOpen
  \bibfield  {author} {\bibinfo {author} {\bibfnamefont {E.~V.}\ \bibnamefont {Castro}}, \bibinfo {author} {\bibfnamefont {K.~S.}\ \bibnamefont {Novoselov}}, \bibinfo {author} {\bibfnamefont {S.~V.}\ \bibnamefont {Morozov}}, \bibinfo {author} {\bibfnamefont {N.~M.}\ \bibnamefont {Peres}}, \bibinfo {author} {\bibfnamefont {J.~M.}\ \bibnamefont {Dos~Santos}}, \bibinfo {author} {\bibfnamefont {J.}~\bibnamefont {Nilsson}}, \bibinfo {author} {\bibfnamefont {F.}~\bibnamefont {Guinea}}, \bibinfo {author} {\bibfnamefont {A.~K.}\ \bibnamefont {Geim}},\ and\ \bibinfo {author} {\bibfnamefont {A.~H.}\ \bibnamefont {Neto}},\ }\bibfield  {title} {\bibinfo {title} {{Biased bilayer graphene: Semiconductor with a gap tunable by the electric field effect}},\ }\href {https://doi.org/10.1103/PHYSREVLETT.99.216802/FIGURES/3/MEDIUM} {\bibfield  {journal} {\bibinfo  {journal} {Physical Review Letters}\ }\textbf {\bibinfo {volume} {99}},\ \bibinfo {pages} {216802} (\bibinfo {year} {2007})}\BibitemShut {NoStop}%
\bibitem [{\citenamefont {Zhang}\ \emph {et~al.}(2009)\citenamefont {Zhang}, \citenamefont {Tang}, \citenamefont {Girit}, \citenamefont {Hao}, \citenamefont {Martin}, \citenamefont {Zettl}, \citenamefont {Crommie}, \citenamefont {Shen},\ and\ \citenamefont {Wang}}]{Zhang2009DirectGraphene.}%
  \BibitemOpen
  \bibfield  {author} {\bibinfo {author} {\bibfnamefont {Y.}~\bibnamefont {Zhang}}, \bibinfo {author} {\bibfnamefont {T.~T.}\ \bibnamefont {Tang}}, \bibinfo {author} {\bibfnamefont {C.}~\bibnamefont {Girit}}, \bibinfo {author} {\bibfnamefont {Z.}~\bibnamefont {Hao}}, \bibinfo {author} {\bibfnamefont {M.~C.}\ \bibnamefont {Martin}}, \bibinfo {author} {\bibfnamefont {A.}~\bibnamefont {Zettl}}, \bibinfo {author} {\bibfnamefont {M.~F.}\ \bibnamefont {Crommie}}, \bibinfo {author} {\bibfnamefont {Y.~R.}\ \bibnamefont {Shen}},\ and\ \bibinfo {author} {\bibfnamefont {F.}~\bibnamefont {Wang}},\ }\bibfield  {title} {\bibinfo {title} {{Direct observation of a widely tunable bandgap in bilayer graphene.}},\ }\href {https://doi.org/10.1038/NATURE08105} {\bibfield  {journal} {\bibinfo  {journal} {Nature}\ }\textbf {\bibinfo {volume} {459}},\ \bibinfo {pages} {820} (\bibinfo {year} {2009})}\BibitemShut {NoStop}%
\bibitem [{\citenamefont {Oostinga}\ \emph {et~al.}(2007)\citenamefont {Oostinga}, \citenamefont {Heersche}, \citenamefont {Liu}, \citenamefont {Morpurgo},\ and\ \citenamefont {Vandersypen}}]{Oostinga2007Gate-inducedDevices}%
  \BibitemOpen
  \bibfield  {author} {\bibinfo {author} {\bibfnamefont {J.~B.}\ \bibnamefont {Oostinga}}, \bibinfo {author} {\bibfnamefont {H.~B.}\ \bibnamefont {Heersche}}, \bibinfo {author} {\bibfnamefont {X.}~\bibnamefont {Liu}}, \bibinfo {author} {\bibfnamefont {A.~F.}\ \bibnamefont {Morpurgo}},\ and\ \bibinfo {author} {\bibfnamefont {L.~M.}\ \bibnamefont {Vandersypen}},\ }\bibfield  {title} {\bibinfo {title} {{Gate-induced insulating state in bilayer graphene devices}},\ }\href {https://doi.org/10.1038/nmat2082} {\bibfield  {journal} {\bibinfo  {journal} {Nature Materials 2007 7:2}\ }\textbf {\bibinfo {volume} {7}},\ \bibinfo {pages} {151} (\bibinfo {year} {2007})}\BibitemShut {NoStop}%
\bibitem [{\citenamefont {Wang}\ \emph {et~al.}(2018)\citenamefont {Wang}, \citenamefont {Chernikov}, \citenamefont {Glazov}, \citenamefont {Heinz}, \citenamefont {Marie}, \citenamefont {Amand},\ and\ \citenamefont {Urbaszek}}]{Wang2018Colloquium:Dichalcogenides}%
  \BibitemOpen
  \bibfield  {author} {\bibinfo {author} {\bibfnamefont {G.}~\bibnamefont {Wang}}, \bibinfo {author} {\bibfnamefont {A.}~\bibnamefont {Chernikov}}, \bibinfo {author} {\bibfnamefont {M.~M.}\ \bibnamefont {Glazov}}, \bibinfo {author} {\bibfnamefont {T.~F.}\ \bibnamefont {Heinz}}, \bibinfo {author} {\bibfnamefont {X.}~\bibnamefont {Marie}}, \bibinfo {author} {\bibfnamefont {T.}~\bibnamefont {Amand}},\ and\ \bibinfo {author} {\bibfnamefont {B.}~\bibnamefont {Urbaszek}},\ }\bibfield  {title} {\bibinfo {title} {{Colloquium: Excitons in atomically thin transition metal dichalcogenides}},\ }\href {https://doi.org/10.1103/REVMODPHYS.90.021001/FIGURES/8/MEDIUM} {\bibfield  {journal} {\bibinfo  {journal} {Reviews of Modern Physics}\ }\textbf {\bibinfo {volume} {90}},\ \bibinfo {pages} {021001} (\bibinfo {year} {2018})}\BibitemShut {NoStop}%
\bibitem [{\citenamefont {Pedersen}(2015)}]{Pedersen2015IntrabandGeneration}%
  \BibitemOpen
  \bibfield  {author} {\bibinfo {author} {\bibfnamefont {T.~G.}\ \bibnamefont {Pedersen}},\ }\bibfield  {title} {\bibinfo {title} {{Intraband effects in excitonic second-harmonic generation}},\ }\href {https://doi.org/10.1103/PHYSREVB.92.235432/FIGURES/4/MEDIUM} {\bibfield  {journal} {\bibinfo  {journal} {Physical Review B - Condensed Matter and Materials Physics}\ }\textbf {\bibinfo {volume} {92}},\ \bibinfo {pages} {235432} (\bibinfo {year} {2015})}\BibitemShut {NoStop}%
\bibitem [{\citenamefont {Aversa}\ and\ \citenamefont {Sipe}(1995)}]{Aversa1995NonlinearAnalysis}%
  \BibitemOpen
  \bibfield  {author} {\bibinfo {author} {\bibfnamefont {C.}~\bibnamefont {Aversa}}\ and\ \bibinfo {author} {\bibfnamefont {J.~E.}\ \bibnamefont {Sipe}},\ }\bibfield  {title} {\bibinfo {title} {{Nonlinear optical susceptibilities of semiconductors: Results with a length-gauge analysis}},\ }\href {https://doi.org/10.1103/PhysRevB.52.14636} {\bibfield  {journal} {\bibinfo  {journal} {Physical Review B}\ }\textbf {\bibinfo {volume} {52}},\ \bibinfo {pages} {14636} (\bibinfo {year} {1995})}\BibitemShut {NoStop}%
\bibitem [{E-P()}]{E-PeriodicaElectrodynamics}%
  \BibitemOpen
  \href {https://www.e-periodica.ch/digbib/view?pid=hpa-001:1952:25::814#421} {\bibinfo {title} {{E-Periodica - On the Definition of the Renormalization Constants in Quantum Electrodynamics}}}\BibitemShut {NoStop}%
\bibitem [{\citenamefont {Lehmann}(1954)}]{Lehmann1954UberFelder}%
  \BibitemOpen
  \bibfield  {author} {\bibinfo {author} {\bibfnamefont {H.}~\bibnamefont {Lehmann}},\ }\bibfield  {title} {\bibinfo {title} {{{\"{U}}ber Eigenschaften von Ausbreitungsfunktionen und Renormierungskonstanten quantisierter Felder}},\ }\href {https://doi.org/10.1007/BF02783624/METRICS} {\bibfield  {journal} {\bibinfo  {journal} {Il Nuovo Cimento}\ }\textbf {\bibinfo {volume} {11}},\ \bibinfo {pages} {342} (\bibinfo {year} {1954})}\BibitemShut {NoStop}%
\bibitem [{\citenamefont {Quintela}\ \emph {et~al.}(2024)\citenamefont {Quintela}, \citenamefont {Peres},\ and\ \citenamefont {Pedersen}}]{Quintela2024TunableGraphene}%
  \BibitemOpen
  \bibfield  {author} {\bibinfo {author} {\bibfnamefont {M.~F. C.~M.}\ \bibnamefont {Quintela}}, \bibinfo {author} {\bibfnamefont {N.~M.~R.}\ \bibnamefont {Peres}},\ and\ \bibinfo {author} {\bibfnamefont {T.~G.}\ \bibnamefont {Pedersen}},\ }\bibfield  {title} {\bibinfo {title} {{Tunable nonlinear excitonic optical response in biased bilayer graphene}},\ }\href {https://doi.org/10.1103/PHYSREVB.110.085433/FIGURES/9/MEDIUM} {\bibfield  {journal} {\bibinfo  {journal} {Physical Review B}\ }\textbf {\bibinfo {volume} {110}},\ \bibinfo {pages} {085433} (\bibinfo {year} {2024})}\BibitemShut {NoStop}%
\bibitem [{\citenamefont {Epstein}\ \emph {et~al.}(2020{\natexlab{b}})\citenamefont {Epstein}, \citenamefont {Chaves}, \citenamefont {Rhodes}, \citenamefont {Frank}, \citenamefont {Watanabe}, \citenamefont {Taniguchi}, \citenamefont {Giessen}, \citenamefont {Hone}, \citenamefont {Peres},\ and\ \citenamefont {Koppens}}]{Epstein2020HighlySemiconductors}%
  \BibitemOpen
  \bibfield  {author} {\bibinfo {author} {\bibfnamefont {I.}~\bibnamefont {Epstein}}, \bibinfo {author} {\bibfnamefont {A.~J.}\ \bibnamefont {Chaves}}, \bibinfo {author} {\bibfnamefont {D.~A.}\ \bibnamefont {Rhodes}}, \bibinfo {author} {\bibfnamefont {B.}~\bibnamefont {Frank}}, \bibinfo {author} {\bibfnamefont {K.}~\bibnamefont {Watanabe}}, \bibinfo {author} {\bibfnamefont {T.}~\bibnamefont {Taniguchi}}, \bibinfo {author} {\bibfnamefont {H.}~\bibnamefont {Giessen}}, \bibinfo {author} {\bibfnamefont {J.~C.}\ \bibnamefont {Hone}}, \bibinfo {author} {\bibfnamefont {N.~M.~R.}\ \bibnamefont {Peres}},\ and\ \bibinfo {author} {\bibfnamefont {F.~H.~L.}\ \bibnamefont {Koppens}},\ }\bibfield  {title} {\bibinfo {title} {{Highly confined in-plane propagating exciton-polaritons on monolayer semiconductors}},\ }\href {https://doi.org/10.1088/2053-1583/AB8DD4} {\bibfield  {journal} {\bibinfo  {journal} {2D Materials}\ }\textbf {\bibinfo {volume} {7}},\ \bibinfo {pages} {035031} (\bibinfo {year}
  {2020}{\natexlab{b}})}\BibitemShut {NoStop}%
\bibitem [{\citenamefont {Eini}\ \emph {et~al.}(2022)\citenamefont {Eini}, \citenamefont {Asherov}, \citenamefont {Mazor},\ and\ \citenamefont {Epstein}}]{Eini2022Valley-polarizedFrequencies}%
  \BibitemOpen
  \bibfield  {author} {\bibinfo {author} {\bibfnamefont {T.}~\bibnamefont {Eini}}, \bibinfo {author} {\bibfnamefont {T.}~\bibnamefont {Asherov}}, \bibinfo {author} {\bibfnamefont {Y.}~\bibnamefont {Mazor}},\ and\ \bibinfo {author} {\bibfnamefont {I.}~\bibnamefont {Epstein}},\ }\bibfield  {title} {\bibinfo {title} {{Valley-polarized hyperbolic exciton polaritons in few-layer two-dimensional semiconductors at visible frequencies}},\ }\href {https://doi.org/10.1103/PhysRevB.106.L201405} {\bibfield  {journal} {\bibinfo  {journal} {Physical Review B}\ }\textbf {\bibinfo {volume} {106}},\ \bibinfo {pages} {L201405} (\bibinfo {year} {2022})}\BibitemShut {NoStop}%
\bibitem [{\citenamefont {Li}\ \emph {et~al.}(2015)\citenamefont {Li}, \citenamefont {Lewin}, \citenamefont {Kretinin}, \citenamefont {Caldwell}, \citenamefont {Novoselov}, \citenamefont {Taniguchi}, \citenamefont {Watanabe}, \citenamefont {Gaussmann},\ and\ \citenamefont {Taubner}}]{Li2015HyperbolicFocusing}%
  \BibitemOpen
  \bibfield  {author} {\bibinfo {author} {\bibfnamefont {P.}~\bibnamefont {Li}}, \bibinfo {author} {\bibfnamefont {M.}~\bibnamefont {Lewin}}, \bibinfo {author} {\bibfnamefont {A.~V.}\ \bibnamefont {Kretinin}}, \bibinfo {author} {\bibfnamefont {J.~D.}\ \bibnamefont {Caldwell}}, \bibinfo {author} {\bibfnamefont {K.~S.}\ \bibnamefont {Novoselov}}, \bibinfo {author} {\bibfnamefont {T.}~\bibnamefont {Taniguchi}}, \bibinfo {author} {\bibfnamefont {K.}~\bibnamefont {Watanabe}}, \bibinfo {author} {\bibfnamefont {F.}~\bibnamefont {Gaussmann}},\ and\ \bibinfo {author} {\bibfnamefont {T.}~\bibnamefont {Taubner}},\ }\bibfield  {title} {\bibinfo {title} {{Hyperbolic phonon-polaritons in boron nitride for near-field optical imaging and focusing}},\ }\href {https://doi.org/10.1038/ncomms8507} {\bibfield  {journal} {\bibinfo  {journal} {Nature Communications 2015 6:1}\ }\textbf {\bibinfo {volume} {6}},\ \bibinfo {pages} {1} (\bibinfo {year} {2015})}\BibitemShut {NoStop}%
\bibitem [{\citenamefont {Kats}\ \emph {et~al.}(2024)\citenamefont {Kats}, \citenamefont {Eini},\ and\ \citenamefont {Epstein}}]{Kats20242DMaterials}%
  \BibitemOpen
  \bibfield  {author} {\bibinfo {author} {\bibfnamefont {I.}~\bibnamefont {Kats}}, \bibinfo {author} {\bibfnamefont {T.}~\bibnamefont {Eini}},\ and\ \bibinfo {author} {\bibfnamefont {I.}~\bibnamefont {Epstein}},\ }\bibfield  {title} {\bibinfo {title} {{2D Semiconductors Superlattices as Hyperbolic Materials}},\ }\href {https://arxiv.org/abs/2411.14785v1} {\bibfield  {journal} {\bibinfo  {journal} {arXiv:2411.14785}\ } (\bibinfo {year} {2024})}\BibitemShut {NoStop}%
\bibitem [{\citenamefont {Woessner}\ \emph {et~al.}(2014)\citenamefont {Woessner}, \citenamefont {Lundeberg}, \citenamefont {Gao}, \citenamefont {Principi}, \citenamefont {Alonso-Gonz{\'{a}}lez}, \citenamefont {Carrega}, \citenamefont {Watanabe}, \citenamefont {Taniguchi}, \citenamefont {Vignale}, \citenamefont {Polini}, \citenamefont {Hone}, \citenamefont {Hillenbrand},\ and\ \citenamefont {Koppens}}]{Woessner2014HighlyHeterostructures}%
  \BibitemOpen
  \bibfield  {author} {\bibinfo {author} {\bibfnamefont {A.}~\bibnamefont {Woessner}}, \bibinfo {author} {\bibfnamefont {M.~B.}\ \bibnamefont {Lundeberg}}, \bibinfo {author} {\bibfnamefont {Y.}~\bibnamefont {Gao}}, \bibinfo {author} {\bibfnamefont {A.}~\bibnamefont {Principi}}, \bibinfo {author} {\bibfnamefont {P.}~\bibnamefont {Alonso-Gonz{\'{a}}lez}}, \bibinfo {author} {\bibfnamefont {M.}~\bibnamefont {Carrega}}, \bibinfo {author} {\bibfnamefont {K.}~\bibnamefont {Watanabe}}, \bibinfo {author} {\bibfnamefont {T.}~\bibnamefont {Taniguchi}}, \bibinfo {author} {\bibfnamefont {G.}~\bibnamefont {Vignale}}, \bibinfo {author} {\bibfnamefont {M.}~\bibnamefont {Polini}}, \bibinfo {author} {\bibfnamefont {J.}~\bibnamefont {Hone}}, \bibinfo {author} {\bibfnamefont {R.}~\bibnamefont {Hillenbrand}},\ and\ \bibinfo {author} {\bibfnamefont {F.~H.~L.}\ \bibnamefont {Koppens}},\ }\bibfield  {title} {\bibinfo {title} {{Highly confined low-loss plasmons in graphene–boron nitride heterostructures}},\ }\href
  {https://doi.org/10.1038/nmat4169} {\bibfield  {journal} {\bibinfo  {journal} {Nature Materials 2014 14:4}\ }\textbf {\bibinfo {volume} {14}},\ \bibinfo {pages} {421} (\bibinfo {year} {2014})}\BibitemShut {NoStop}%
\bibitem [{\citenamefont {Lundeberg}\ \emph {et~al.}(2017)\citenamefont {Lundeberg}, \citenamefont {Gao}, \citenamefont {Asgari}, \citenamefont {Tan}, \citenamefont {Duppen}, \citenamefont {Autore}, \citenamefont {Alonso-Gonz{\'{a}}lez}, \citenamefont {Woessner}, \citenamefont {Watanabe}, \citenamefont {Taniguchi}, \citenamefont {Hillenbrand}, \citenamefont {Hone}, \citenamefont {Polini},\ and\ \citenamefont {Koppens}}]{Lundeberg2017TuningPlasmonics}%
  \BibitemOpen
  \bibfield  {author} {\bibinfo {author} {\bibfnamefont {M.~B.}\ \bibnamefont {Lundeberg}}, \bibinfo {author} {\bibfnamefont {Y.}~\bibnamefont {Gao}}, \bibinfo {author} {\bibfnamefont {R.}~\bibnamefont {Asgari}}, \bibinfo {author} {\bibfnamefont {C.}~\bibnamefont {Tan}}, \bibinfo {author} {\bibfnamefont {B.~V.}\ \bibnamefont {Duppen}}, \bibinfo {author} {\bibfnamefont {M.}~\bibnamefont {Autore}}, \bibinfo {author} {\bibfnamefont {P.}~\bibnamefont {Alonso-Gonz{\'{a}}lez}}, \bibinfo {author} {\bibfnamefont {A.}~\bibnamefont {Woessner}}, \bibinfo {author} {\bibfnamefont {K.}~\bibnamefont {Watanabe}}, \bibinfo {author} {\bibfnamefont {T.}~\bibnamefont {Taniguchi}}, \bibinfo {author} {\bibfnamefont {R.}~\bibnamefont {Hillenbrand}}, \bibinfo {author} {\bibfnamefont {J.}~\bibnamefont {Hone}}, \bibinfo {author} {\bibfnamefont {M.}~\bibnamefont {Polini}},\ and\ \bibinfo {author} {\bibfnamefont {F.~H.}\ \bibnamefont {Koppens}},\ }\bibfield  {title} {\bibinfo {title} {{Tuning quantum nonlocal effects in graphene
  plasmonics}},\ }\href {https://doi.org/10.1126/SCIENCE.AAN2735} {\bibfield  {journal} {\bibinfo  {journal} {Science}\ }\textbf {\bibinfo {volume} {357}},\ \bibinfo {pages} {187} (\bibinfo {year} {2017})}\BibitemShut {NoStop}%
\bibitem [{\citenamefont {Iranzo}\ \emph {et~al.}(2018)\citenamefont {Iranzo}, \citenamefont {Nanot}, \citenamefont {Dias}, \citenamefont {Epstein}, \citenamefont {Peng}, \citenamefont {Efetov}, \citenamefont {Lundeberg}, \citenamefont {Parret}, \citenamefont {Osmond}, \citenamefont {Hong}, \citenamefont {Kong}, \citenamefont {Englund}, \citenamefont {Peres},\ and\ \citenamefont {Koppens}}]{Iranzo2018ProbingHeterostructure}%
  \BibitemOpen
  \bibfield  {author} {\bibinfo {author} {\bibfnamefont {D.~A.}\ \bibnamefont {Iranzo}}, \bibinfo {author} {\bibfnamefont {S.}~\bibnamefont {Nanot}}, \bibinfo {author} {\bibfnamefont {E.~J.}\ \bibnamefont {Dias}}, \bibinfo {author} {\bibfnamefont {I.}~\bibnamefont {Epstein}}, \bibinfo {author} {\bibfnamefont {C.}~\bibnamefont {Peng}}, \bibinfo {author} {\bibfnamefont {D.~K.}\ \bibnamefont {Efetov}}, \bibinfo {author} {\bibfnamefont {M.~B.}\ \bibnamefont {Lundeberg}}, \bibinfo {author} {\bibfnamefont {R.}~\bibnamefont {Parret}}, \bibinfo {author} {\bibfnamefont {J.}~\bibnamefont {Osmond}}, \bibinfo {author} {\bibfnamefont {J.~Y.}\ \bibnamefont {Hong}}, \bibinfo {author} {\bibfnamefont {J.}~\bibnamefont {Kong}}, \bibinfo {author} {\bibfnamefont {D.~R.}\ \bibnamefont {Englund}}, \bibinfo {author} {\bibfnamefont {N.~M.}\ \bibnamefont {Peres}},\ and\ \bibinfo {author} {\bibfnamefont {F.~H.}\ \bibnamefont {Koppens}},\ }\bibfield  {title} {\bibinfo {title} {{Probing the ultimate plasmon confinement limits with a van
  der Waals heterostructure}},\ }\href {https://doi.org/10.1126/SCIENCE.AAR8438} {\bibfield  {journal} {\bibinfo  {journal} {Science}\ }\textbf {\bibinfo {volume} {360}},\ \bibinfo {pages} {291} (\bibinfo {year} {2018})}\BibitemShut {NoStop}%
\bibitem [{\citenamefont {Epstein}\ \emph {et~al.}(2020{\natexlab{c}})\citenamefont {Epstein}, \citenamefont {Alcaraz}, \citenamefont {Huang}, \citenamefont {Pusapati}, \citenamefont {Hugonin}, \citenamefont {Kumar}, \citenamefont {Deputy}, \citenamefont {Khodkov}, \citenamefont {Rappoport}, \citenamefont {Hong}, \citenamefont {Peres}, \citenamefont {Kong}, \citenamefont {Smith},\ and\ \citenamefont {Koppens}}]{Epstein2020Far-fieldVolumes}%
  \BibitemOpen
  \bibfield  {author} {\bibinfo {author} {\bibfnamefont {I.}~\bibnamefont {Epstein}}, \bibinfo {author} {\bibfnamefont {D.}~\bibnamefont {Alcaraz}}, \bibinfo {author} {\bibfnamefont {Z.}~\bibnamefont {Huang}}, \bibinfo {author} {\bibfnamefont {V.~V.}\ \bibnamefont {Pusapati}}, \bibinfo {author} {\bibfnamefont {J.~P.}\ \bibnamefont {Hugonin}}, \bibinfo {author} {\bibfnamefont {A.}~\bibnamefont {Kumar}}, \bibinfo {author} {\bibfnamefont {X.~M.}\ \bibnamefont {Deputy}}, \bibinfo {author} {\bibfnamefont {T.}~\bibnamefont {Khodkov}}, \bibinfo {author} {\bibfnamefont {T.~G.}\ \bibnamefont {Rappoport}}, \bibinfo {author} {\bibfnamefont {J.~Y.}\ \bibnamefont {Hong}}, \bibinfo {author} {\bibfnamefont {N.~M.}\ \bibnamefont {Peres}}, \bibinfo {author} {\bibfnamefont {J.}~\bibnamefont {Kong}}, \bibinfo {author} {\bibfnamefont {D.~R.}\ \bibnamefont {Smith}},\ and\ \bibinfo {author} {\bibfnamefont {F.~H.}\ \bibnamefont {Koppens}},\ }\bibfield  {title} {\bibinfo {title} {{Far-field excitation of single graphene plasmon
  cavities with ultracompressed mode volumes}},\ }\href {https://doi.org/10.1126/SCIENCE.ABB1570} {\bibfield  {journal} {\bibinfo  {journal} {Science}\ }\textbf {\bibinfo {volume} {368}},\ \bibinfo {pages} {1219} (\bibinfo {year} {2020}{\natexlab{c}})}\BibitemShut {NoStop}%
\bibitem [{\citenamefont {Menabde}\ \emph {et~al.}(2021)\citenamefont {Menabde}, \citenamefont {Lee}, \citenamefont {Lee}, \citenamefont {Ha}, \citenamefont {Heiden}, \citenamefont {Yoo}, \citenamefont {Kim}, \citenamefont {Low}, \citenamefont {Lee}, \citenamefont {Oh},\ and\ \citenamefont {Jang}}]{Menabde2021Real-spaceDeposition}%
  \BibitemOpen
  \bibfield  {author} {\bibinfo {author} {\bibfnamefont {S.~G.}\ \bibnamefont {Menabde}}, \bibinfo {author} {\bibfnamefont {I.~H.}\ \bibnamefont {Lee}}, \bibinfo {author} {\bibfnamefont {S.}~\bibnamefont {Lee}}, \bibinfo {author} {\bibfnamefont {H.}~\bibnamefont {Ha}}, \bibinfo {author} {\bibfnamefont {J.~T.}\ \bibnamefont {Heiden}}, \bibinfo {author} {\bibfnamefont {D.}~\bibnamefont {Yoo}}, \bibinfo {author} {\bibfnamefont {T.~T.}\ \bibnamefont {Kim}}, \bibinfo {author} {\bibfnamefont {T.}~\bibnamefont {Low}}, \bibinfo {author} {\bibfnamefont {Y.~H.}\ \bibnamefont {Lee}}, \bibinfo {author} {\bibfnamefont {S.~H.}\ \bibnamefont {Oh}},\ and\ \bibinfo {author} {\bibfnamefont {M.~S.}\ \bibnamefont {Jang}},\ }\bibfield  {title} {\bibinfo {title} {{Real-space imaging of acoustic plasmons in large-area graphene grown by chemical vapor deposition}},\ }\href {https://doi.org/10.1038/s41467-021-21193-5} {\bibfield  {journal} {\bibinfo  {journal} {Nature Communications 2021 12:1}\ }\textbf {\bibinfo {volume} {12}},\
  \bibinfo {pages} {1} (\bibinfo {year} {2021})}\BibitemShut {NoStop}%
\bibitem [{\citenamefont {Stern}(1967)}]{Stern1967PolarizabilityGas}%
  \BibitemOpen
  \bibfield  {author} {\bibinfo {author} {\bibfnamefont {F.}~\bibnamefont {Stern}},\ }\bibfield  {title} {\bibinfo {title} {{Polarizability of a Two-Dimensional Electron Gas}},\ }\href {https://doi.org/10.1103/PhysRevLett.18.546} {\bibfield  {journal} {\bibinfo  {journal} {Physical Review Letters}\ }\textbf {\bibinfo {volume} {18}},\ \bibinfo {pages} {546} (\bibinfo {year} {1967})}\BibitemShut {NoStop}%
\bibitem [{\citenamefont {Agarwal}\ \emph {et~al.}(1974)\citenamefont {Agarwal}, \citenamefont {Pattanayak},\ and\ \citenamefont {Wolf}}]{Agarwal1974ElectromagneticMedia}%
  \BibitemOpen
  \bibfield  {author} {\bibinfo {author} {\bibfnamefont {G.~S.}\ \bibnamefont {Agarwal}}, \bibinfo {author} {\bibfnamefont {D.~N.}\ \bibnamefont {Pattanayak}},\ and\ \bibinfo {author} {\bibfnamefont {E.}~\bibnamefont {Wolf}},\ }\bibfield  {title} {\bibinfo {title} {{Electromagnetic fields in spatially dispersive media}},\ }\href {https://doi.org/10.1103/PhysRevB.10.1447} {\bibfield  {journal} {\bibinfo  {journal} {Physical Review B}\ }\textbf {\bibinfo {volume} {10}},\ \bibinfo {pages} {1447} (\bibinfo {year} {1974})}\BibitemShut {NoStop}%
\bibitem [{\citenamefont {Hopfield}\ and\ \citenamefont {Thomas}(1963)}]{Hopfield1963TheoreticalCrystals}%
  \BibitemOpen
  \bibfield  {author} {\bibinfo {author} {\bibfnamefont {J.~J.}\ \bibnamefont {Hopfield}}\ and\ \bibinfo {author} {\bibfnamefont {D.~G.}\ \bibnamefont {Thomas}},\ }\bibfield  {title} {\bibinfo {title} {{Theoretical and Experimental Effects of Spatial Dispersion on the Optical Properties of Crystals}},\ }\href {https://doi.org/10.1103/PhysRev.132.563} {\bibfield  {journal} {\bibinfo  {journal} {Physical Review}\ }\textbf {\bibinfo {volume} {132}},\ \bibinfo {pages} {563} (\bibinfo {year} {1963})}\BibitemShut {NoStop}%
\bibitem [{\citenamefont {Gershuni}\ and\ \citenamefont {Epstein}(2024)}]{Gershuni2024In-planeLoss}%
  \BibitemOpen
  \bibfield  {author} {\bibinfo {author} {\bibfnamefont {Y.}~\bibnamefont {Gershuni}}\ and\ \bibinfo {author} {\bibfnamefont {I.}~\bibnamefont {Epstein}},\ }\bibfield  {title} {\bibinfo {title} {{In-plane exciton polaritons versus plasmon polaritons: Nonlocal corrections, confinement, and loss}},\ }\bibfield  {journal} {\bibinfo  {journal} {Physical Review B}\ }\textbf {\bibinfo {volume} {109}},\ \href {https://doi.org/10.1103/PHYSREVB.109.L121408} {10.1103/PHYSREVB.109.L121408} (\bibinfo {year} {2024})\BibitemShut {NoStop}%
\end{thebibliography}%

\end{document}